\documentclass[12pt,a4paper]{article}

\usepackage{amsmath,amssymb,amsfonts}
\usepackage[dvips]{graphicx}
\usepackage{epsfig}
\usepackage{subcaption}
\usepackage[T1]{fontenc}
\usepackage{amsmath}
\usepackage{amsfonts}
\usepackage{amssymb}
\usepackage{color}
\usepackage{graphicx}
\usepackage{caption}
\usepackage{subcaption}
\usepackage{comment}%}
\usepackage{cite}
\usepackage{amsmath,amssymb,amsfonts,graphicx}
\usepackage{epsfig}

 \newcommand{\be}{\begin{equation}}
	 \newcommand{\ee}{\end{equation}}
	 \newcommand{\ba}{\begin{eqnarray}}
		 \newcommand{\ea}{\end{eqnarray}}
		 
		   \newcommand{\bea}{\begin{eqnarray}}
			 \newcommand{\eea}{\end{eqnarray}}

\newcommand{\beq}{\begin{eqnarray}}
\newcommand{\eeq}{\end{eqnarray}}

\usepackage[left=2.4cm,top=3.3cm,right=2.4cm,bottom=3.3cm,bindingoffset=0cm]{geometry}

\begin{document}

\title{ \begin{flushright}\ \vskip -2.5cm {\footnotesize {IFUP-TH-2019}}\end{flushright}
\vskip 30pt
Vortons with Abelian and non-Abelian currents \\ and their stability}

\vskip 30pt 

\author{Gianni Tallarita$^a$, Adam Peterson$^b$, \\ Stefano Bolognesi$^c$, and Peter Bedford$^b$  \\ 
\vspace{0 cm}\\
\normalsize \it $^{a}$Departamento de Ciencias, Facultad de Artes Liberales, Universidad Adolfo Ib\'a\~nez,\\
\normalsize \it Santiago 7941169, Chile.
 \vspace{.3 cm}\\
\normalsize \it $^{b}$University of Toronto, Department of Physics, Toronto, ON M5S 1A7, Canada.
 \vspace{.3 cm}\\
\normalsize \it $^{c}$Department of Physics ``E. Fermi'', University of Pisa and INFN, Sezione di Pisa\\{\normalsize \it Largo Pontecorvo, 3, Ed. C, 56127 Pisa, Italy}
 \vspace{.4 cm}\\
{ \footnotesize gianni.tallarita@uai.cl, apeterson@utoronto.ca, stefano.bolognesi@unipi.it, peter.bedford@mail.utoronto.ca}  
}

\date{\hfill}

\maketitle

\begin{abstract}

We explore vorton solutions in the  Witten's $U(1) \times U(1)$ model for cosmic strings and in a  modified version $U(1) \times SO(3)$ obtained by introducing a triplet of non-Abelian fields to condense inside the string. 
We restrict to the case in which the unbroken symmetry in the bulk remains global.
The vorton solutions are found  numerically for certain choices of parameters and compared with an analytical solutions obtained in the thin vorton limit.  
%{\bf ... aggiungere ...}
%We consider a stability analysis of vortons to perturbations of the $SO(3)$ field.
We also discuss the vorton decay into Q-rings (or spinning Q-balls) and, to some extent, the time dependent behavior of vortons above the charge threshold.

\end{abstract}

\newpage

\tableofcontents

\section{Introduction}

 Vortons are closed loops of superconducting vortex strings which are made dynamically stable by a current flowing around the circle. This creates a  balance of angular momentum  against the string tension.  
Originally, cosmic vortons were introduced in \cite{Davis:1988ij} following the work by Witten \cite{Witten:1984eb}.  
They have since become a topic of interest to physicists due to their complex structure and stability characteristics (see \cite{Volovik:2003fe} for example).  %%% Added a reference to G. Volovik.  %%%
They have important cosmological implications in early universe, in any model admitting superconducting cosmic  strings, for example in the grand unified theory scenarios  \cite{Brandenberger:1996zp} and in the production of baryon asymmetry \cite{Davis:1996zg,Masperi:1998ct}.  Vortons 
%resolve the charge-monopole s-wave scattering ambiguity \cite{Davisother}, 
can be sources for high energy cosmic rays \cite{Masperi:1997qf} and are even relevant in the construction of dark matter models \cite{Martins:1999jg}.  In general thay can be relavant in any setup, from QCD to condensed matter, where there are vortices with some internal degrees of freedom.

In fact, it was only relatively recently that vortons were constructed and analyzed numerically in a global version of Witten's original $U(1) \times U(1)$ model \cite{Radu:2008pp}.  More recently they have been constructed in the gauged versions, although their stability in these models remains an elusive topic due to the time dependent gauge dynamics required for a proper analysis \cite{Garaud:2013iba}. The general concept of vorton stability has long been a difficult and interesting topic \cite{Lemperiere:2002en}\,-\!\cite{Battye:2008mm} and to this day a complete classification of vorton stability is lacking, even in the simplest global model of  \cite{Radu:2008pp}.

The first purpose of this work is to make progress in understanding the spectrum and stability of vorton solutions. We will concentrate  only on models for which there are no massless gauge fields in the bulk. 
Among the many questions one would like to understand there are: how the mass of the vorton is related to the conserved charges, what is the range of stability of vortons and what they decay into if they are not stable. We will answer, partially,   these questions by using two different strategies, one numerical and one analytical. The analytical approach is done in the thin vorton limit.  We can perform the two approaches in different regions of the parameter space, so the comparison between the two can only be qualitative.

Another purpose of this paper is to explore vorton with strings that possess additional degrees of freedom following from the condensation of non-Abelian fields. 
We consider a modified version of the global  Witten model that has symmetry group $U(1) \times SO(3) $
 and a field transforming under $SO(3)$ that condenses inside the string.
This type of non-Abelian orientational moduli  on vortices has been discussed previously in various contexts  in \cite{Shifman:2012vv}\,-\!\cite{Tallarita:2019czh}.  
For the particular model we consider, vorton solutions can be constructed from solutions in the original $U(1) \times U(1)$ model by mapping the global $U(1)$ to a subgroup of $SO(3)$ and the $S^1$ internal moduli space to a $S^1 \subset S^2$ at a certain latitude of the sphere.  In this case the momentum from a conserved $SO(3)$ current running along the vorton loop leads to a dynamic stability of the vorton string.  While we present analytical suggestions that solutions at different latitudes should occur, at least in some particular limits, we will see that numerical construction is complicated and only for the equator ones we have been able to find explicit solutions.

%%% Revision request #1: The section numbers have been corrected to reflect the correct sections in the paper. %%%
We will organize the paper as follows.
In Section \ref{due} we will review the $U(1) \times U(1)$  model and  introduce the modified $U(1) \times SO(3)$ model describing the general ansatz for vortons solutions.  
In Section \ref{tre} we will explore the thin vorton limit.
In Section \ref{num} we provide the  various numerical solutions.
We will conclude with a discussion of results in Section \ref{cinque}.

\section{Vortons in the Abelian and non-Abelian model}
\label{due}

We begin our discussion with a brief review of the Witten $U(1) \times U(1)$ model \cite{Witten:1984eb}.  
We will discuss the version of this model with one $U(1)$ considered as a gauge group, and the other $U(1)$ group as global.
The Lagrangian is the following:
%%% Revision request #2: Some additional information is included to clarify the role of each field, including the gauge fields.  %%%
\begin{equation}
\mathcal{L} = -\frac{1}{4}F_{\mu\nu}F^{\mu\nu}+D_\mu \phi^{\star}D^\mu \phi + \partial_\mu \sigma^{\star}\partial^\mu \sigma-U(|\phi|,|\sigma|) \ ,
\label{WittenModel}
\end{equation}
where the gauge covariant derivative is given by
\begin{equation}
D_\mu \phi = (\partial_\mu + {\rm i}g A_\mu)\phi \ .
\end{equation}
The model has a $U(1)$ gauge symmetry under which $\phi$ is charged, and a $U(1)$ global symmetry under which $\sigma$ is charged.  Here $A_\mu$ is the gauge field with field strength $F_{\mu \nu}$ for the $U(1)$ gauge symmetry. 
The potential is
\begin{equation}
U(|\phi|, |\sigma|) = \frac{1}{4} \lambda_\phi (|\phi|^2 - \eta_\phi^2)^2+\frac{1}{4} \lambda_\sigma|\sigma|^2 (|\sigma|^2 - 2 \eta_\sigma^2 ) + \gamma |\phi|^2|\sigma|^2 \ ,
\end{equation} 
where $\lambda_\phi$ , $\lambda_\sigma$, $\eta_\phi$, $\eta_\sigma$, and $\gamma$ are constants. 
This is the most generic potential, up to quartic interactions, consistent with the symmetries of the problem. 
The equations of motion follow from the  action principle, and can be stated as follows:
\bea
D_\mu D^\mu \phi + \frac{\partial U}{\partial\phi^*}&=&0 \ , \nonumber \\
\partial_\mu \partial^\mu \sigma + \frac{\partial U}{\partial \sigma^*} &=&0 \ , \nonumber \\
\phantom{\frac{U}{U}} \partial_\mu F^{\mu\nu} &=& g j^\nu \ , 
\label{EOM}
\eea
where the gauge $U(1)$ current is given by
\begin{equation}
j^\mu = {\rm i}\left( \phi^{\star} D^\mu \phi - \phi D^\mu \phi^{\star}\right) \ .
\end{equation}
The field $\phi$  condenses in the vacuum $|\phi| = \eta_\phi$, therefore the gauge $U(1)$ is broken and the theory admits topological vortices carring magnetic flux.  The global $U(1)$ is unbroken in the vacuum $\sigma = 0$.
In general, when $\gamma$ is sufficiently large compared to $\frac{1}{2}\lambda_\sigma \eta_\sigma^2$ the field $\sigma$ will condense in the core of the vortex. This breaks the $U(1)$ symmetry in the core, thus giving  massless moduli  localized on the vortex with target space a circle $S^1$.

Vortons are, roughly speaking, solutions to the equations of motion (\ref{EOM}) that describe $U(1)$ vortices that are bent around into a  ring.  In order to counteract the inward tension the ring must have an angular momentum providing an outward force.  The angular momentum is given by a current of the condensate $\sigma$ spinning around the loop.
Some vorton  solutions in this model have been previously constructed numerically in  \cite{Radu:2008pp}.  We describe our procedure for solving them here. 

%%% Revision request #3: A brief explanation of the ansatz is included to clarify the logic of its selection. %%%
Working in cylindrical coordinates $(t,r,z,\varphi)$ we write our ansatz as:
\begin{align}
&\phi = f_1(r,z) \,{e}^{{\rm i} \psi(r,z)}\ , \nonumber \\
&\sigma = f_2(r,z) \, { e}^{{\rm i}m\varphi +{\rm i}\omega t}\ ,
\label{InitialAnsatz}
\end{align}
where $m$ is an integer.  Briefly, the profile $f_1(r,z)$ and the phase $\psi(r,z)$ are the standard profile and phase functions of vortex solutions that are modified to represent a vortex ring wrapped around the $z$-axis. In order to represent a spinning vortex with topological charge $n$ the phase function $\psi(r,z)$ must wind $n$ times around the large contour $C$ representing the boundary of the half-plane $r \ge 0$. Actually, in this articular gauge, the winding will happen all in the axes $r=0$.  Regularity requires $f_1(r,z)$ to have $n$ zeros within the boundary contour.
For non-zero $m$ the solution has a $\varphi$ dependence with a non-zero angular momentum.  Regularity for non-zero $m$ requires $f_2(r,z)$ to vanish along the line $r = 0$.  It is easy to show that this ansatz is a valid solution, consistent with the equations of motion (\ref{EOM}).  A more complete explanation of this ansatz is given in \cite{Radu:2008pp}.

%%% Revision request #4:  A reference to the equations of motion below has been added to clarify that the assumptions between eqs (7) and (8) are consistent.  %%%
For the gauge field components we  work in the gauge $A_r = 0$
\begin{equation}
A = A_\mu dx^\mu = A_0 dt + A_z dz + A_\varphi d\varphi \ .
\end{equation}
By writing out the equations of motion for $A_\mu$, it is possible to show that the components $A_0$ and $A_\varphi$ can be consistently assumed to be vanishing, as shown in (\ref{FinalForm}) below.  Thus we further reduce the ansatz to
\begin{equation}
A = A_z(r,z) dz \ .
\end{equation}

%%% Revision request #5: The following paragraph is edited to clarify the motivation.  %%%
For numerically constructing solutions it is easiest to recast the form of the ansatz to simplify the form of the equations of motion.  We write:
\begin{align}
&\phi(r,z) = X(r,z) + {\rm i} Y(r,z) \nonumber \\
&\sigma(r,z) = Z(r,z){ e}^{{\rm i}m\varphi +{\rm i}\omega t} \ . 
\end{align}
From the form of (\ref{InitialAnsatz}) we then have:
\begin{align}
&X(r,z) = f_1 \cos \psi \ , \nonumber \\
&Y(r,z) = f_1 \sin \psi\ , \nonumber \\
&Z(r,z) = f_2\ .
\end{align}
With this form of the ansatz we find the equations to solve:
\bea
\nabla^2 X -g(\partial_z A_z Y+2 A_z \partial_z Y)- g^2A_z^2 X - \left( \frac{\lambda_\phi}{2}\left(X^2 +Y^2-1\right) + \gamma Z^2\right) X &=& 0\ , \nonumber \\
\nabla^2 Y+g(\partial_z A_z X+2 A_z \partial_z X)- g^2A_z^2 Y - \left( \frac{\lambda_\phi}{2}\left(X^2 +Y^2-1\right) + \gamma Z^2\right) Y &=& 0\ , \nonumber \\
\nabla^2 Z - \left( \frac{m^2}{r^2} - \omega^2 +\frac{\lambda_\sigma}{2}\left(Z^2 - \eta_\sigma^2 \right) + \gamma \left( X^2 + Y^2\right)\right) Z &=& 0\ , \nonumber \\
\partial_r^2 A_z +\frac{1}{r} \partial_r A_z - 2g(Y\partial_z X-X\partial_z Y)-2g^2 (X^2+Y^2)A_z &=& 0 \ .
\label{FinalForm}
\eea
We will solve this system using a relaxation procedure following from the energy minimization condition.   However, the procedure cannot be implemented directly due to the requirement of charge conservation.  To achieve this we solve the system using a Legendre transformation to hold the unbroken $U(1)$ charge $Q$ constant.
The unbroken $U(1)$ Noether charge $Q = i \int d^3x(  \sigma\dot{\sigma^*}-\sigma^*\dot{\sigma} )$ is given by
\begin{equation}
Q = 2 \omega \mathcal{N} \ ,
\label{Charge1}
\end{equation}
where
\begin{equation}
\mathcal{N} = \int d^3x Z^2\ .
\label{N1}
\end{equation} %%% Revision request #6: The reference has been corrected.  %%%
At each iteration of the relaxation procedure, the value of $\omega$ is calculated from (\ref{Charge1}) and substituted back into (\ref{FinalForm}). This ensures our numerical procedure will solve minimising the energy functional at constant $Q$.

%%% Revision request #7:  Although we do not wish to include the full details of the ground state solutions, we have added a reference to previous work where this analysis was carried out.  It is straight forward but tedious and distracting to include here.  %%%
We now modify the previous model  by replacing the $U(1)$ field $\sigma$ with a non-Abelian $SO(3)$ field $\chi^i$. We are therefore interested in finding stable vorton rings in the following model:
\begin{equation}
\mathcal{L} = -\frac{1}{4}F_{\mu\nu}F^{\mu\nu}+D_\mu \phi^{\star}D^\mu \phi + \partial_\mu \chi^i\partial^\mu \chi^i-U(|\phi|,|\chi|) \ ,
\label{Model}
\end{equation}
with potential terms
\begin{equation}
U(|\phi|, |\chi|) = \frac{\lambda_\phi }{4} (|\phi|^2 - \eta_\phi^2)^2+\frac{\lambda_\chi}{4}\chi^i\chi^i (\chi^i\chi^i - 2 \eta_\chi^2 ) + \gamma |\phi|^2\chi^i\chi^i \ .
\end{equation}
We have chosen the numerical values for the constants such that vortices within the model develop non-zero profiles for the $\chi^i$ field in their cores.  In other words the $SO(3)$ symmetry is required to be broken in vortex cores and unbroken in the vacuum.  A detailed analysis of the various vacuum states can be found in \cite{Peterson:2014nma}.

Note that  the solutions to (\ref{Model}) are the same as in the Witten model (\ref{WittenModel}) when the non-Abelian field $\chi$ is chosen to be constrained to, for example,  the internal $i = \{1,2\}$ plane $\chi^3=0$.  What is novel about this model is that the choice of the internal plane is arbitrary and moreover a third component for the field $\chi$  might in principle develop.
Here we assume the ansatz:
\begin{align}
&\chi^1 = Z_1(r,z) \cos(m \varphi + \omega t) \nonumber \\
&\chi^2 = Z_1(r,z) \sin(m \varphi + \omega t) \nonumber \\
&\chi^3 = Z_2 (r,z),
\label{NAAnsatz}
\end{align}
with the ansatz for the fields $\phi$ and $A_z$ considered as in the Abelian model above. %%% Revision request #8: We clarify that the gauge field is as considered above. %%%
For the non-Abelian case, thhe unbroken $U(1) \subset SO(3)$ Noether charge $Q_i =  i \int d^3x \epsilon_{ijk}\chi_j\dot{\chi_k}$ is given by
\begin{equation}
Q_3 = 2 \omega \mathcal{N}_3 \ ,
\label{Charge2}
\end{equation}
where
\begin{equation}
\mathcal{N}_3 = \int d^3x Z_1^2\ .
\label{N2}
\end{equation}
When we expand the equations of motion assuming the form (\ref{NAAnsatz}) we find:  %%% Revision request #9: The typos in the equation above have been corrected. They are now consistent.%%%
\bea
\nabla^2 X -g(\partial_z A_z Y+2 A_z \partial_z Y)- g^2A_z^2 X - \left( \frac{\lambda_\phi}{2}\left(X^2 +Y^2-1\right) + \gamma (Z_1^2+Z_2^2)\right) X &= & 0 \ , \nonumber \\
\nabla^2 Y+g(\partial_z A_z X+2 A_z \partial_z X)- g^2A_z^2 Y - \left( \frac{\lambda_\phi}{2}\left(X^2 +Y^2-1\right) + \gamma (Z_1^2+Z_2^2)\right) Y &= & 0 \ , \nonumber \\
\nabla^2 Z_1 - \left( \frac{m^2}{r^2} - \omega^2 +\frac{\lambda_\sigma}{2}\left(Z_1^2+Z_2^2 - \eta_\chi^2 \right) + \gamma \left( X^2 + Y^2\right)\right) Z_1 &= & 0 \ , \nonumber \\
\nabla^2 Z_2 - \left(\frac{\lambda_\sigma}{2}\left(Z_1^2+Z_2^2 - \eta_\chi^2 \right) + \gamma \left( X^2 + Y^2\right)\right) Z_2 &= & 0 \ , \nonumber \\
\partial_r^2 A_z +\frac{1}{r} \partial_r A_z - 2g(Y\partial_z X-X\partial_z Y)-2g^2 (X^2+Y^2)A_z &= & 0 \ .  \nonumber \\
\label{FinalForm2}
\eea

We apply the following boundary conditions (which are also applied in the Abelian case for the fields $X$, $Y$, $Z = Z_1$ and $A_z$): %%% Revision request #10: It is clarified that the Abelian components are applied with the same boundary conditions above.  We choose not to state the Abelian case explicitly since they are the same and we wish to reduce clutter. %%%
\begin{align}
\partial_r X = \partial_r Y = Z_1 = \partial_r Z_2 = A_z = 0 \ ,\quad {\rm at } \quad r = 0  \ .
\label{BC1}
\end{align}
We also use the symmetry of the solution to constrain the region to $z \ge 0$.  This leads to the boundary conditions:
\begin{equation}
\partial_z X = Y = \partial_z Z_1 = \partial_z Z_2 =  \partial_z A_z = 0 \ ,\quad {\rm at } \quad  z = 0 \ .
\label{BC2}
\end{equation}
Finally, we insist that our solution approach the vacuum values at large distance from the vorton:
\begin{align}
& X \rightarrow 1 \nonumber \\
& Y = Z_1 = Z_2 = A_z \rightarrow 0  \ , \quad \mbox{ as }  \quad \sqrt{r^2 + z^2} \rightarrow \infty \ .
\label{BC3}
\end{align}
%%% Revision request #11: We clarify which specific equations we are referring to in the following paragraph.  %%%
We will discuss numerical solutions of the equations (\ref{FinalForm2}) with boundary conditions (\ref{BC1}), (\ref{BC2}), and (\ref{BC3}) as well as solutions to the Abelian case (\ref{FinalForm}) in Section \ref{num}. Since the solutions come from similar equations, we chose not to display the solutions separately.
Before discussing numerical solutions we wish to present some analytical results in some interesting physical limits.

\section{Thin vorton limit}
\label{tre}

%%% Revision: Several changes to notation have been made in the first paragraphs to simplify the energy momentum tensor formulas, and other quantities.  %%%

In this section we discuss some analytical results which can be derived taking the limit in which the radius of the vorton is much larger than the radius of the constituent vortex; we call this the thin vorton limit (see also \cite{Carter:1990sm}). 
In the thin-vorton limit we can use an effective Lagrangian approach to study the vorton.
The internal moduli is a complex field $S=S^1+ iS^2$ with $|S|^2 = v^2$. Here $v$ will represent the expectation value of $|\vec{S}|$ that is determined by the full potential and kinetic terms in the underlying model in which we are considering a thin vorton approximation. The vorton is a string loop with radius $R$ and action. %%% Revision request #12:  The role of $\mu$ and $v$ are clarified.  %%%
\beq
\label{fixed}
S_{\rm eff} =   \int dt dl  \, \Big( \partial_{\mu} S^{\star} \partial^{\mu} S - T  \Big) \ ,  %%% Correction:  The action is corrected to reflect the proper designation for the tension T.  %%%
\label{eqq}
\eeq
where the component index $\mu =  (t, l)$ and $l: (0,2\pi R]$ and  $T$ is the vortex tension. The vorton ansatz consists in  a winding of the $S$ field around the circle. 
We thus take
\beq
S = v \ e^{i m \phi + i \omega t} \ .
\eeq
Note that this is a solution to on the string derived from (\ref{eqq}).  %%% Revision request #13: We explain the source of the equation of motion following from the string action above. %%%
 Here, $m$ is an integer.  It is the number of windings of $S$ around the third direction. The frequency is 
\beq
\label{ll}
\omega = \frac{m}{ R} \ .\eeq
The vorton is not stabilized by any topological charge. It is instead stabilized by the angular momentum in the third direction. %%%Revision request #14: Formulas for the angular momentum and charge have been added.  %%%
\beq
J_3 = \int dl \; T^0_\varphi = 4 \pi R^2 \omega^2 v^2    = 4 \pi m^2 v^2   \ ,
\eeq
where the energy momentum tensor is given by:
\beq
T_{\mu\nu} = \partial_\mu S^{\star} \partial_\nu S + \partial_\nu S^{\star} \partial_\mu S - \eta_{\mu\nu} \mathcal{L} \ . 
\eeq
This angular momentum is generated by the current circulating around the vorton loop.
There is also a conserved charge $Q$ corresponding to the rotations around the internal circle
\beq
Q  =  i\int dl \; (\partial_t S^{\star} S - S^{\star} \partial_t S ) =  4 \pi m v^2   = \frac{J_3}{m} \ .
\eeq
We write the energy for the vorton
\beq
E =   4 \pi R \omega^2 v^2 +   2 \pi R T  \ ,
\eeq
and then  rewrite the same energy but this time as function of the conserved quantity $J_3$
\beq
\label{efun}
E(R;J_3) =  \frac{ J_3 }{ R}   +   2 \pi R T  \ .
\eeq
This is the quantity that  has to be minimized while keeping $J_3$ fixed, thus obtaining
\beq
\label{remin}
R =\sqrt{ \frac{ J_3}{2 \pi T}}  \ ,  \qquad E =  \sqrt{8 \pi T J_3} \ .
\label{regge1}
\eeq
Since $E$ is the mass $M$ of the vorton we thus have the familiar Regge-trajectory relation between mass-energy and angular momentum $M^2 \propto J_3$. In particular the thin vorton approximation is valid for $m$ sufficiently big such that. %%% Revision request #15:  The relation between mass energy and angular momentum is clarified.  %%%
\beq
R  \gg r \ ,
\eeq
where $r$ is the vortex radius, thus the angular momentum  must be sufficiently large.

\begin{figure}
\centering
\includegraphics[width=0.6\linewidth]{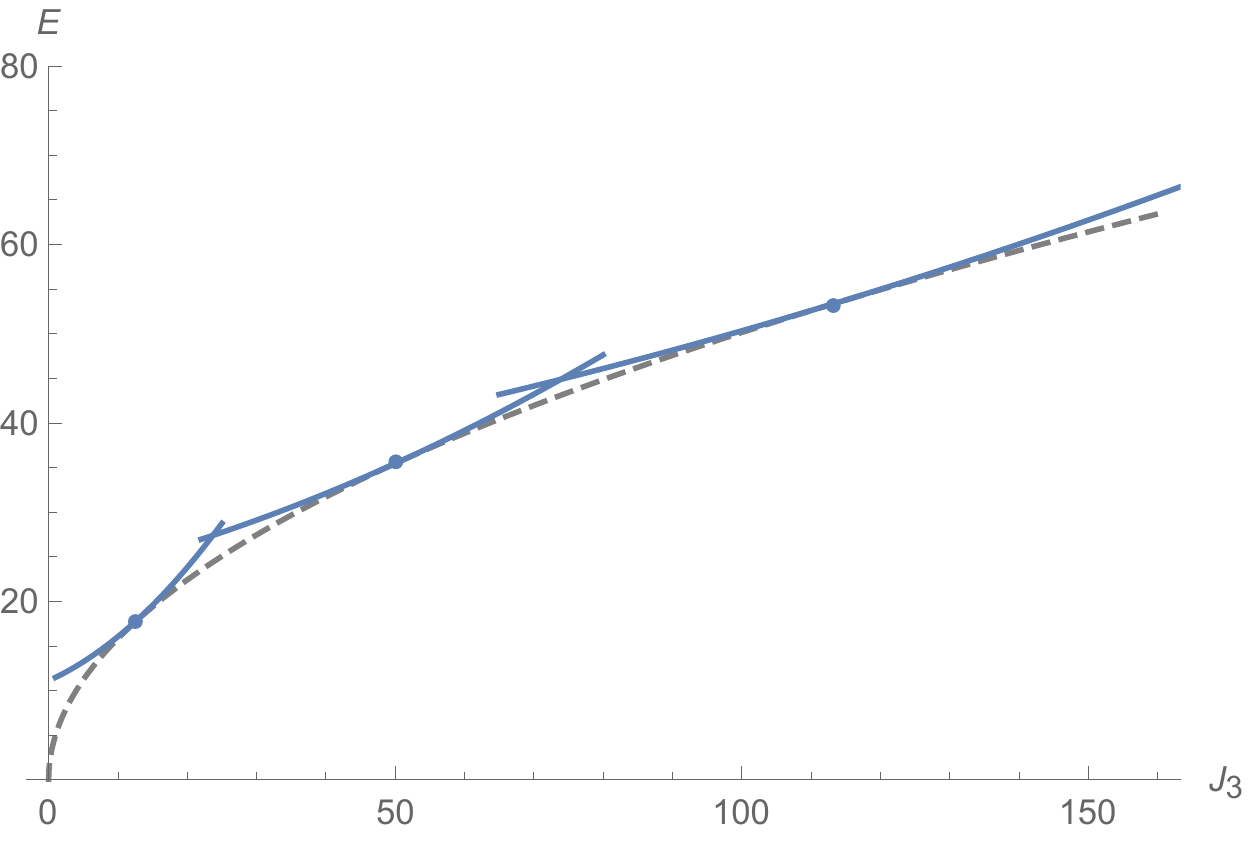}
\caption{Energy vs. angular momentum  for vortons in the case of large $\lambda$; specifically we used here $T=1$, $v=1$ and we plot the values (\ref{enl}) for three cases $m=1,2,3$. The dashed line correspond to the Regge trajectory while the dots correspond to the optimal values when $s=v$.}
\label{regge}
\end{figure}

Generically, when $S$ assumes its vacuum values we have that $|S|^2=v^2$, given the charge $Q$ fixed we have only a discrete set of solutions labeled by the integer $m$. In this case the behaviour of the field on the string worldsheet is light-like  (\ref{ll}). However, this can in general be relaxed if we allow $S$ to deviate from its minimum.
We thus  consider a potential for $|S|$: 
\beq
S_{\rm eff} =   \int dt dl \left( \partial_{\mu} S^{\star} \cdot \partial^{\mu} S - \frac{\lambda^2}{2} (|S|^2-v^2)^2 - T  \right) \ .
\eeq
In the limit $\lambda \to \infty$ we have $|S| \rightarrow v$ and we recover exactly the previous model (\ref{eqq}).  %%%Revision request #16: We explain how the previous model is recovered as \lambda -> infinity. %%%
The solution for the vorton is
\beq
S= s \ e^{i m \phi + i \omega t} \ .
\eeq
where $s$ is not necessarily equal to the minimum $v$;  the equality $s=v$ happens only for a particular vorton, the one for which we have ``optimal'' or light-like conditions (\ref{ll}). Values above or below the optimal condition are sometimes called ``electric'' or ``magnetic'' \cite{Carter:1990sm}.
The equation of motion on the vortex gives
\beq
\omega^2 - \frac{m^2}{R^2} - \lambda^2 (s^2-v^2)s = 0  \ , \label{ome}
\eeq
%"For small deviations from the minimum  $v$ we have
%\beq
%\omega \simeq  \frac{m}{R} + \frac{R \lambda^2 v^2}{m} (s-v) \ .
%\eeq
and the angular momentum in the third direction is given by
\beq
J_3 = 4 \pi R \omega m s^2     \  . \label{jt}
\eeq
From the last two equations we get $\omega$ and $s$ as funtion of the constant $v$ and the conserved quantity $J_3$ and $m$ (or $Q=J_3 / m$ alternatively).
% so we get
%\beq
%s^2 \simeq \frac{J_3}{4 \pi m^2} \ .
%\eeq
We then write the energy for the vorton
\beq
E =   2 \pi R \left( \omega(m,J_3)^2s^2 + \frac{m^2}{R^2}s(m,J_3)^2 +  \frac{\lambda^2}{2} (s(m,J_3)^2-v^2)^2 \right) +   2 \pi R T  \ ,
\eeq
and find the minimum with respect to $R$.
In the limit of small deviation from the minimum of the potential $v$, which we expect to be true in the limit of large $\lambda$, we can expand 
\beq
J_3 = 4 \pi  m^2 v^2 + j \ ,
\label{am}
\eeq
and solve (\ref{ome}) and (\ref{jt}) at various orders in powers of $j$.  In particular we find
\beq
s= v +\frac{ j}{4 \pi  v \left(2 m^2+\lambda ^2 R^2 v^2\right)} + \dots \ ,
\eeq
and 
\beq
E = 2 \pi R T +  \frac{4 \pi  m^2 v^2}{ R} + \frac{ j }{ R}     +  \frac{\lambda^2 R ( 2 m^2   + R^2 v^4 \lambda^2)}{8 \pi m^2 (2 m^2 + R^2 v^3 \lambda^2)^2}j^2    + \dots \ .
\eeq
%and rewriting as function of $J_3$ we find
%\beq
%E(R;J_3) \simeq  2 \pi R T+  \frac{ J_3 }{ R}     +  \frac{\lambda^2 R ( 2 m^2   + R^2 v^4 \lambda^2)}{8 \pi m^2 (2 m^2 %+ R^2 v^3 \lambda^2)^2}(J_3 - 4\pi m^2 v^2)^2    \,
%\eeq
This has to be minimized while keeping $j$ fixed. We find at leading order
\beq
R  =    m v \sqrt{\frac{2}{T}}  + \frac{ j}{4 \pi m v \sqrt{2 T}} +\dots \ ,
\eeq
and
\beq
 E =4 \pi  m \sqrt{2 T} v +\frac{j \sqrt{T}}{\sqrt{2} m v} + v \frac{j^2 \sqrt{2 T} \left(-T^2-2 \lambda ^2 T v^5+2 \lambda ^2 T v^4+\lambda ^4 v^{10}\right)}{32 \pi  m^3 v^3 \left(T+\lambda ^2 v^5\right)^2} + \dots \ .
\label{en}
\eeq
In this case we therefore have a continuum of solutions;  for any given $m$ we can choose an arbitrary $J_3$. 
At the ``optimal'' value $J_3 = 4\pi m^2 v^2$ ($j=0$) we recover the previous results (\ref{regge1}).  
The presence of this potential modifies the Regge trajectories $E\propto J_3 ^{1/2}$ even in the thin vorton limit. 
When $\lambda \to \infty$ the coefficient of $j^2$ goes to a constant while all the other terms in the expansion (not showed explicitly in the formula (\ref{en})) go to zero,  we thus obtain:
\beq
 E =4 \pi  m \sqrt{2 T} v +\frac{j \sqrt{T}}{\sqrt{2} m v} +  \frac{j^2 \sqrt{2 T} }{32 \pi  m^3 v^2} \ .
\label{enl}
\eeq
So the result (\ref{regge1}) is recovered but nevertheless we see a continuum of solutions since a very small deviation from the bottom of the potential can provide a significant variation of the angular momentum. In Figure \ref{regge} we show the energy vs. angular momentum in the case of large $\lambda$ for three cases $m=1,2,3$ and compare it with the Regge trajectory. 
\begin{figure}
\centering
\includegraphics[width=0.6\linewidth]{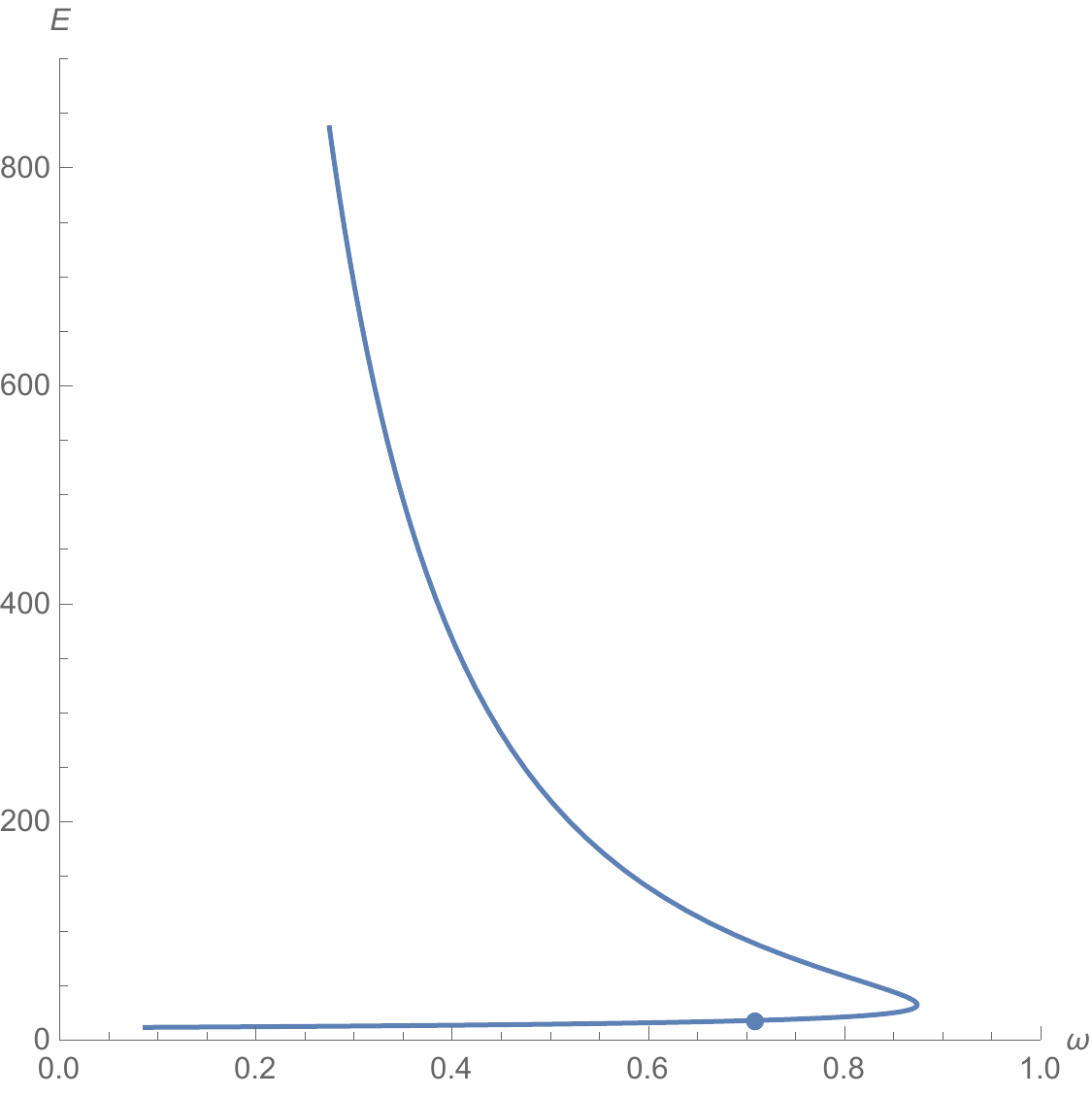}
\caption{Energy vs. $\omega$  for vortons in the case of large $\lambda$; specifically we used here $m=1$, $T=1$, $v=1$.  Note the existence of a critical $\omega$. The dot corresponds to the optimal value where $s=v$.}
\label{vsomega}
\end{figure}

The relation between $J$ and $\omega$ is nonlinear. In fact, if we plot the energy vs. $\omega$ we see a more complex behaviour (see Figure \ref{vsomega}). In particular, there is a maximal value of omega above which no solution exists. For infinite $\lambda$ this can be computed as 
\beq
\omega_{\rm crit} =    \frac{2 \sqrt{2 T}}{(1 + \sqrt{5}) v} \ .
\eeq

If the vortex is non-Abelian the internal moduli are represented by a vector field  $\vec{S}=(S^1,S^2,S^3)$ with $\vec{S}\cdot \vec{S} = v^2$ and the effective action is the same as (\ref{fixed}).
The vorton ansatz involves a winding of the $\vec{S}$ field around a circle in the $S^2$ manifold. We thus take
\beq
S^3 = \cos \alpha \ , \qquad S^1 + i S^2 = \sin{\alpha} \ e^{i m \phi + i \omega t} \ .
\eeq
Note that this is solution to the equation of motion on the string.
The angle $\alpha$ is measuring the latitude of the circle; in general this is not a maximal circle and $\alpha$ will be determined by the charges. Again, $m$ is an integer and it is the number of windings of $ S^1 + i S^2$  around the third direction. Again, the frequency is given in (\ref{ll}).
The angular momentum in the third direction is
\beq
J_3 = 4 \pi R^2 \omega^2 v^2  \sin^2 \alpha  = 4 \pi m^2 v^2 \sin^2 \alpha \ .
\eeq
There is also a conserved charge $Q_3$ corresponding to the rotations around the internal third direction which is related to the angolar momentum and $m$:
\beq
Q_3  =   4 \pi m v^2  \sin^2 \alpha = \frac{J_3}{m} \ .
\eeq
We then write the energy for the vorton as
\beq
E =   4 \pi R \omega^2 v^2 \sin^2 \alpha  +   2 \pi R T \  . 
\eeq
Now we rewrite the same energy, but this time as function of the conserved quantity $J_3$, and we obtain the same as (\ref{efun}) with results from the minimization given by (\ref{remin}).
So  the angle $\alpha$ is determined from the angular momentum as 
\beq
 \sin \alpha  =  \frac{ 1}{2m v} \sqrt{\frac{ J_3 }{\pi }  } \ .
\eeq
Note that it is the quantity $m \sin \alpha$ which is determined by $J_3$. We thus have an exact degeneracy, for a given value of $J_3$ we have multiple vortons labeled by $m$ which have same radius and mass but different internal structure. This  exact degeneracy is most likely an artifact of the approximations we are using.

\section{Numerical solutions and results}
\label{num}

In this section we present numerical solutions to equations (\ref{FinalForm2}). 
To solve for specific vorton solutions we considered axially symmetric solutions on a two dimensional grid in the $(r,z)$-plane. Depending on the details of the solutions we typically considered a grid of $Nr \times Nz$ grid of $100 \times 100$ to $400 \times 400$ on a domain of $Lr \times Lz = 100 \times 100$. Derivatives were calculated using finite difference. To solve the equations we made use of a relaxation procedure through gradient descent. Starting with an initial ansatz represented by $f_0$ satisfying the topological numbers we consider the following iterative procedure:
\begin{equation}
f_{n+1} = f_n + cF\left[f_n\right], 
\label{GradientDescentMethod}
\end{equation}
where $F[f_n]$ are the equations of motion (\ref{FinalForm2}), and $c$ is the learning parameter.  We choose $c$ to achieve convergence in the procedure (typically $c \sim 0.001-0.1$). We continue the procedure until convergence is observed.   Our criterion for convergence is that the norm of the equations of motion on the grid satisfies:
\begin{equation}
\left\lvert\left\lvert\sum F[f_n] \right\rvert\right\rvert \leq 10^{-5},
\end{equation} 
where the sum is over the grid points in the domain.
%%%Revision request #18: The specific criteria for convergence is added and clarified. %%%

We also mention that it is possible to give $c$ a functional form depending on say the radial coordinate $r$, say $c \propto \tanh(r)$.  The purpose of this modification is to improve convergence behavior around the coordinate singularity at $r = 0$. This allows us to choose larger functional values of $c$.  We have checked that this does not affect the stability properties of our solutions which we will discuss below.  Merely the time required to converge is improved.  

Since we consider vortons with a constant charge $Q$, at each step of the iterative procedure we solve for $\omega$ using (\ref{Charge1}) or (\ref{Charge2}), where we integrate (\ref{N1}) or (\ref{N2}) numerically with the current solutions $f_n$.  %%%Revision request #19: It is clarified that either form for Q and N may be used depending on the system considered (either U(1) or SO(3).  In any case they are equivalent. %%%

\begin{figure}[h!]. %%% Revision request #23:  Figures have been enlarged to make axes labels more visible. %%%
\begin{subfigure}{.5\textwidth}
\centering
\includegraphics[width=0.9\linewidth]{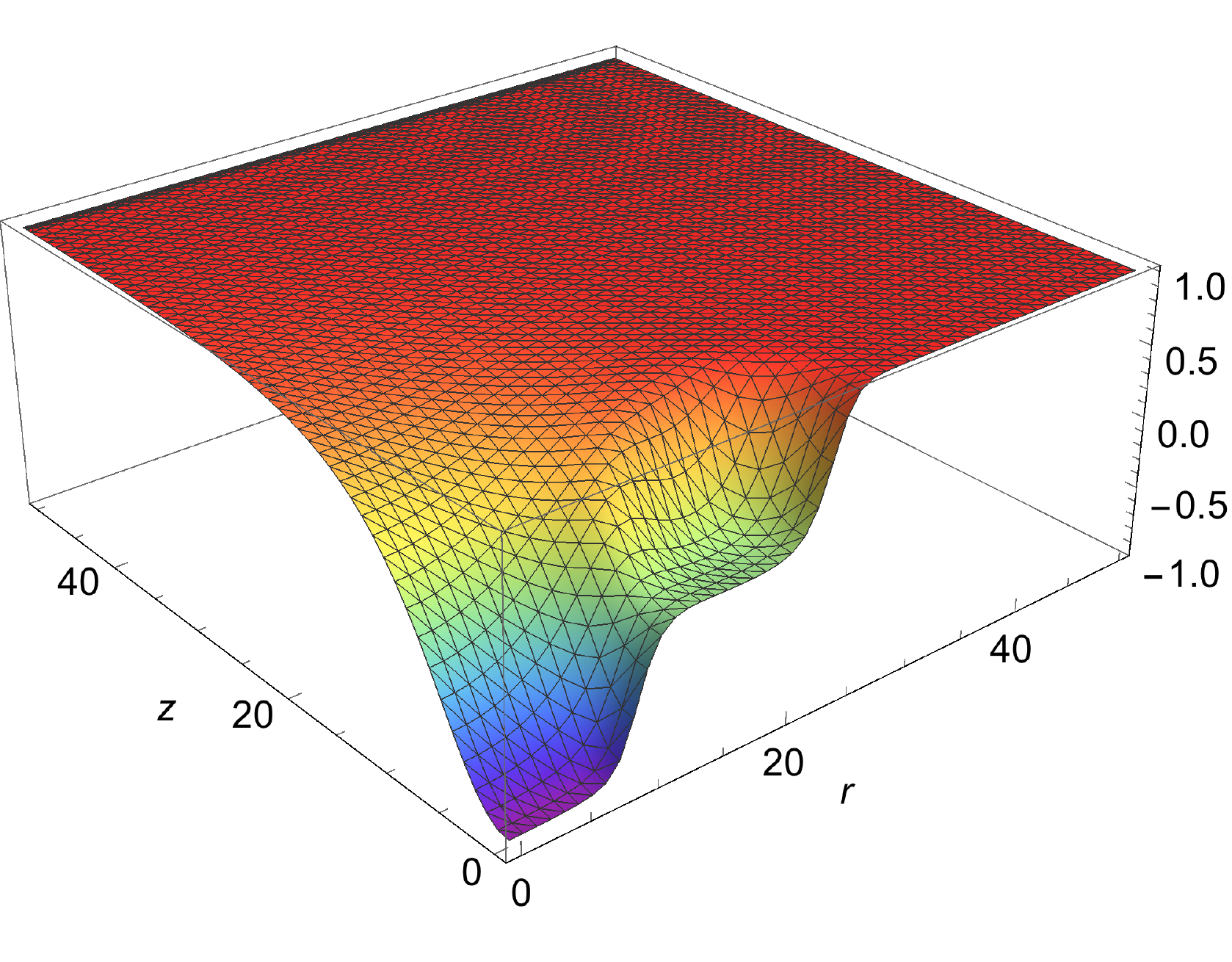}
\caption{$X$ profile}
\end{subfigure}
\begin{subfigure}{.5\textwidth}
\centering
\includegraphics[width=0.9\linewidth]{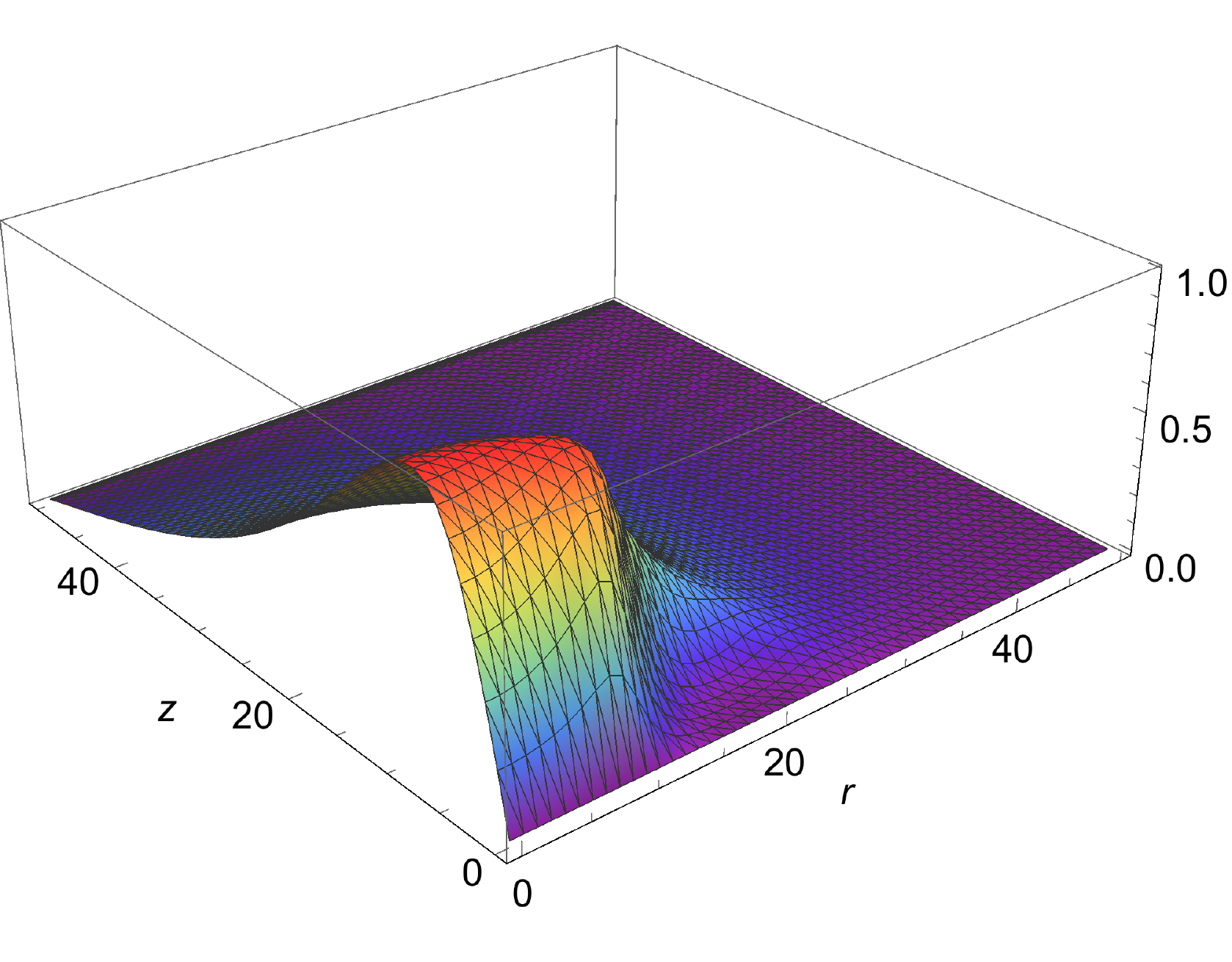}
\caption{$Y$ profile}
\end{subfigure}
\begin{subfigure}{.5\textwidth}
\centering
\includegraphics[width=0.9\linewidth]{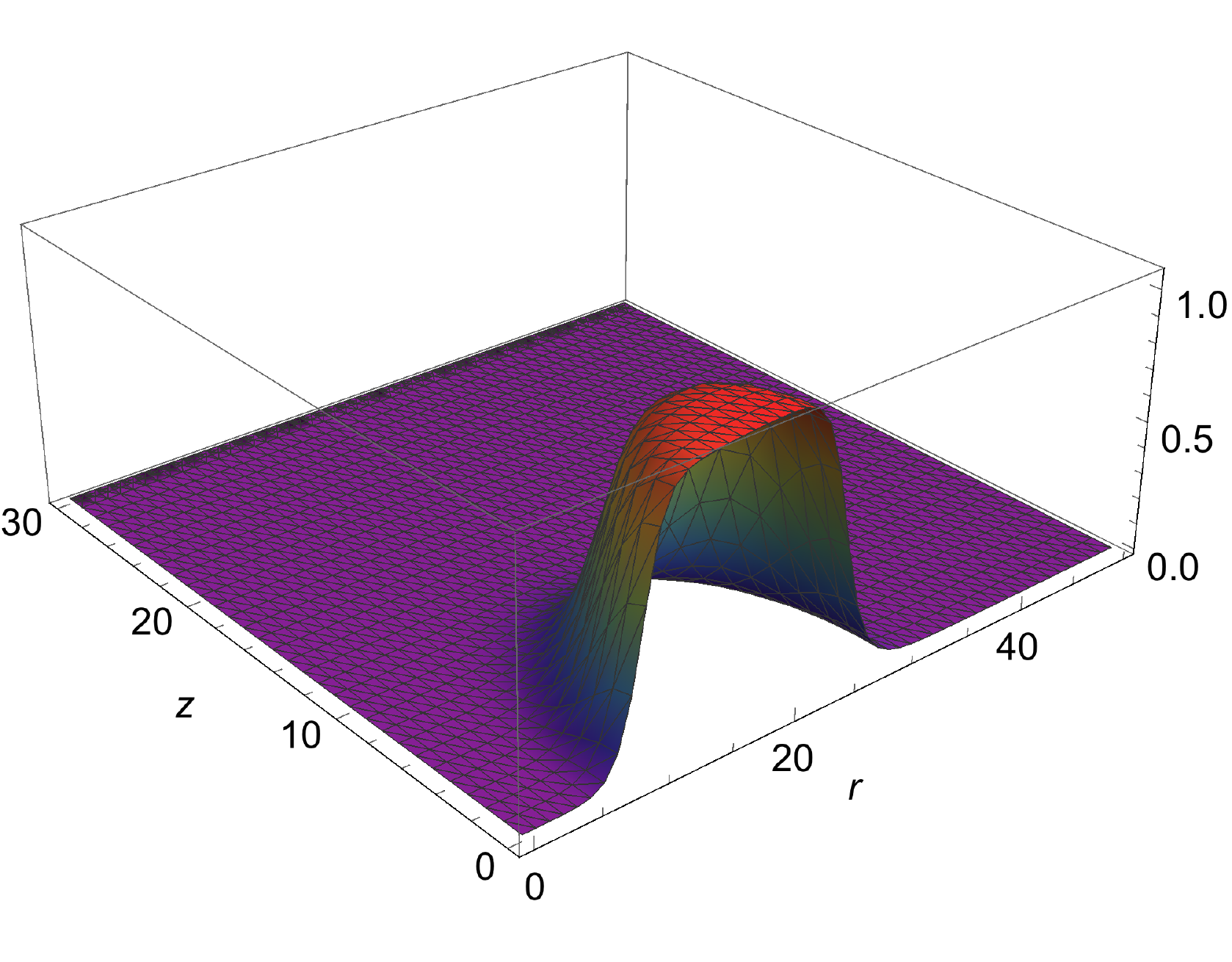}
\caption{$Z_1$ profile}
\end{subfigure}
\begin{subfigure}{.5\textwidth}
\centering
\includegraphics[width=0.9\linewidth]{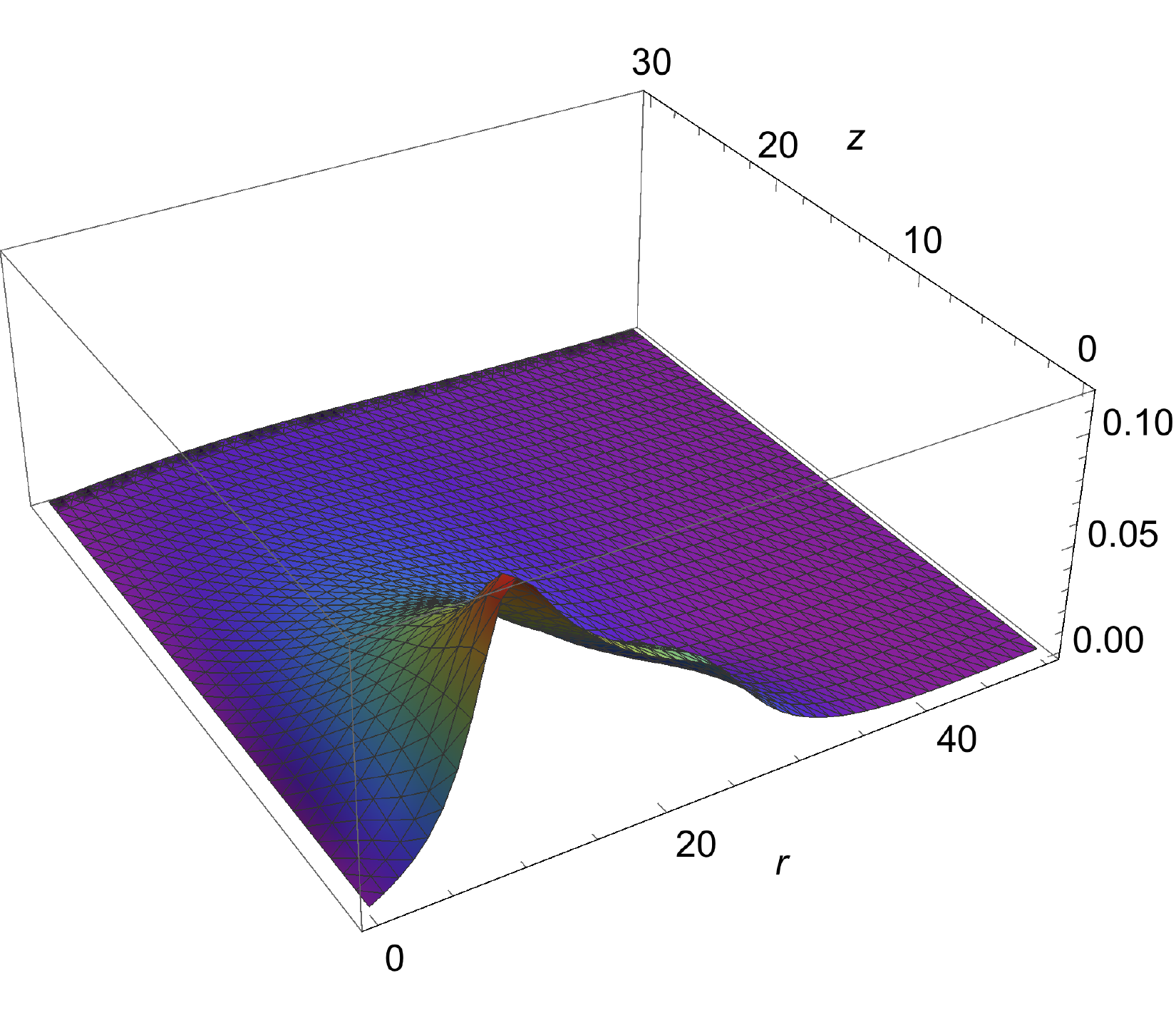}
\caption{$|\vec{B}|$}
\end{subfigure}
\begin{subfigure}{.5\textwidth}
\centering
\includegraphics[width=0.9\linewidth]{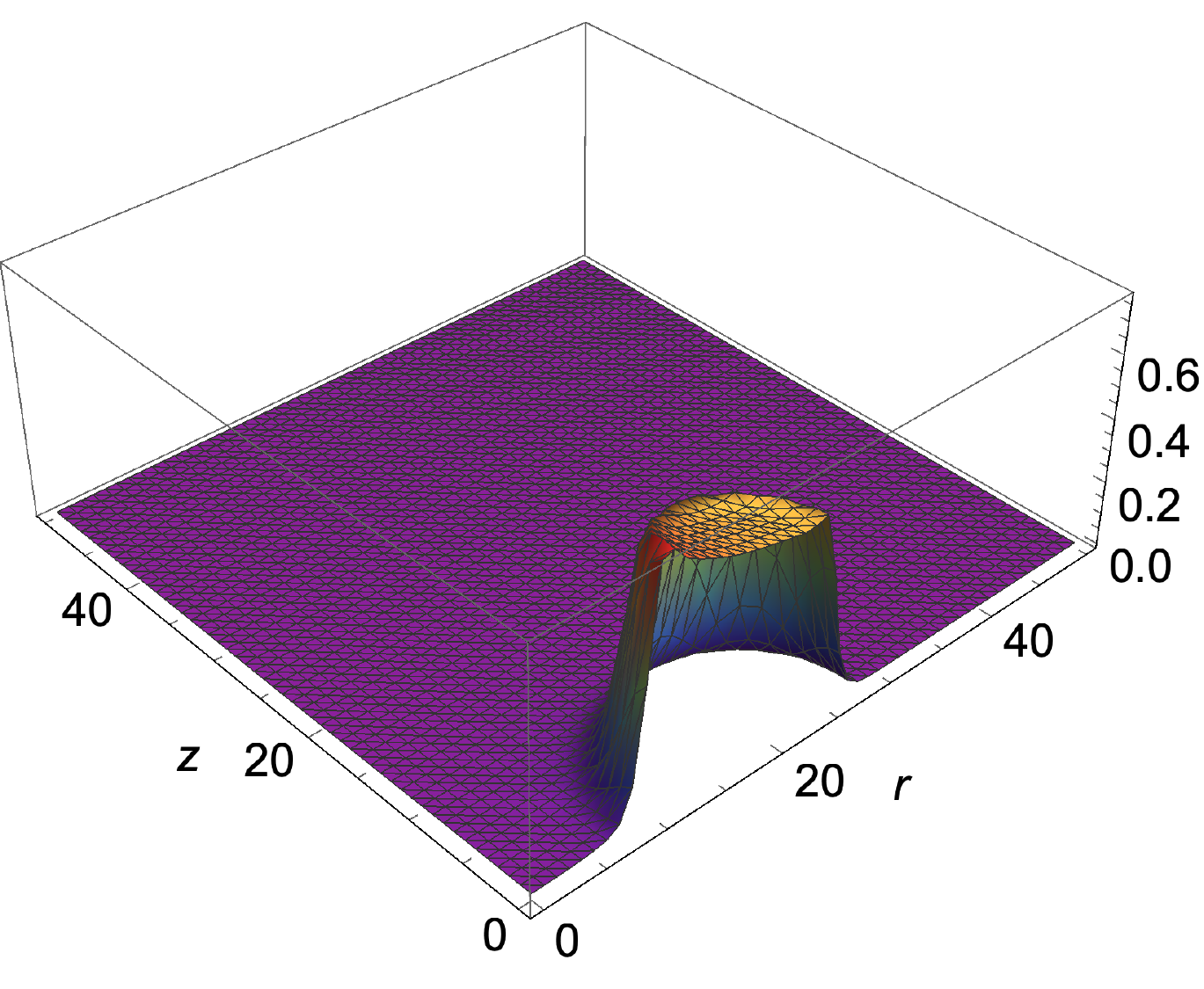}
\caption{Energy density profile}
\end{subfigure}
\caption{Vorton profiles for $X$, $Y$, and $Z_1$ components.  The magnetic field $|\vec{B}|=B_\phi = \partial_r A_z$ and energy density for the solution are also shown.  The parameters are $n=1$, $m = 6$, $\lambda_\phi = 4.5$, $\lambda_\sigma = 4.0$, $\gamma = 2.8$, $Q = 20000$, $g = 0.1$.  For this value of $Q$, we find $\omega = 0.576$.}%
\label{fig1}%
\end{figure}

\begin{figure}[h!].  %%% Revision request #23:  Figures have been enlarged to make axes labels more visible. %%%
\begin{subfigure}{.5\textwidth}
\centering
\includegraphics[width=0.9\linewidth]{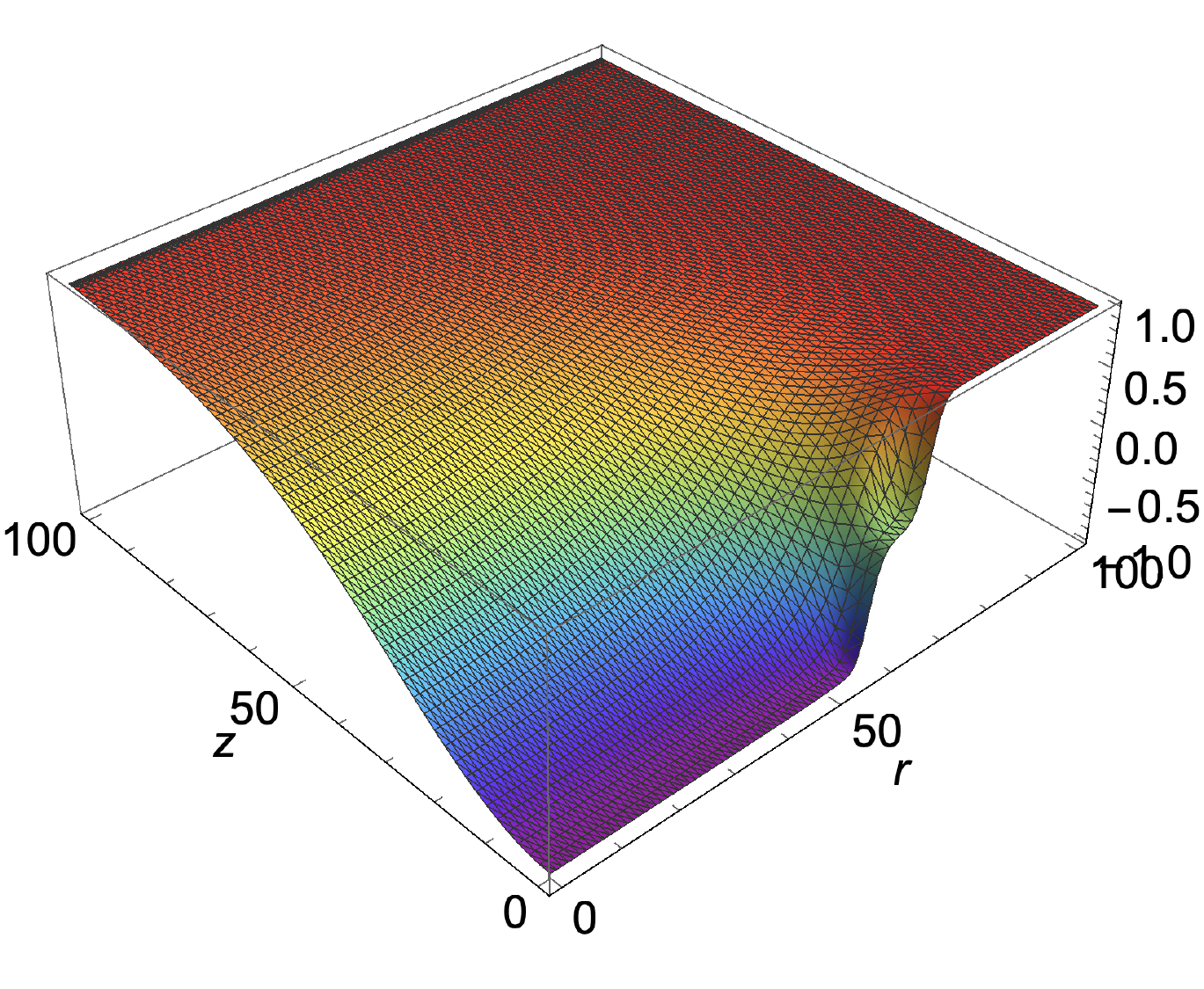}
\caption{$X$ profile}
\end{subfigure}
\begin{subfigure}{.5\textwidth}
\centering
\includegraphics[width=0.9\linewidth]{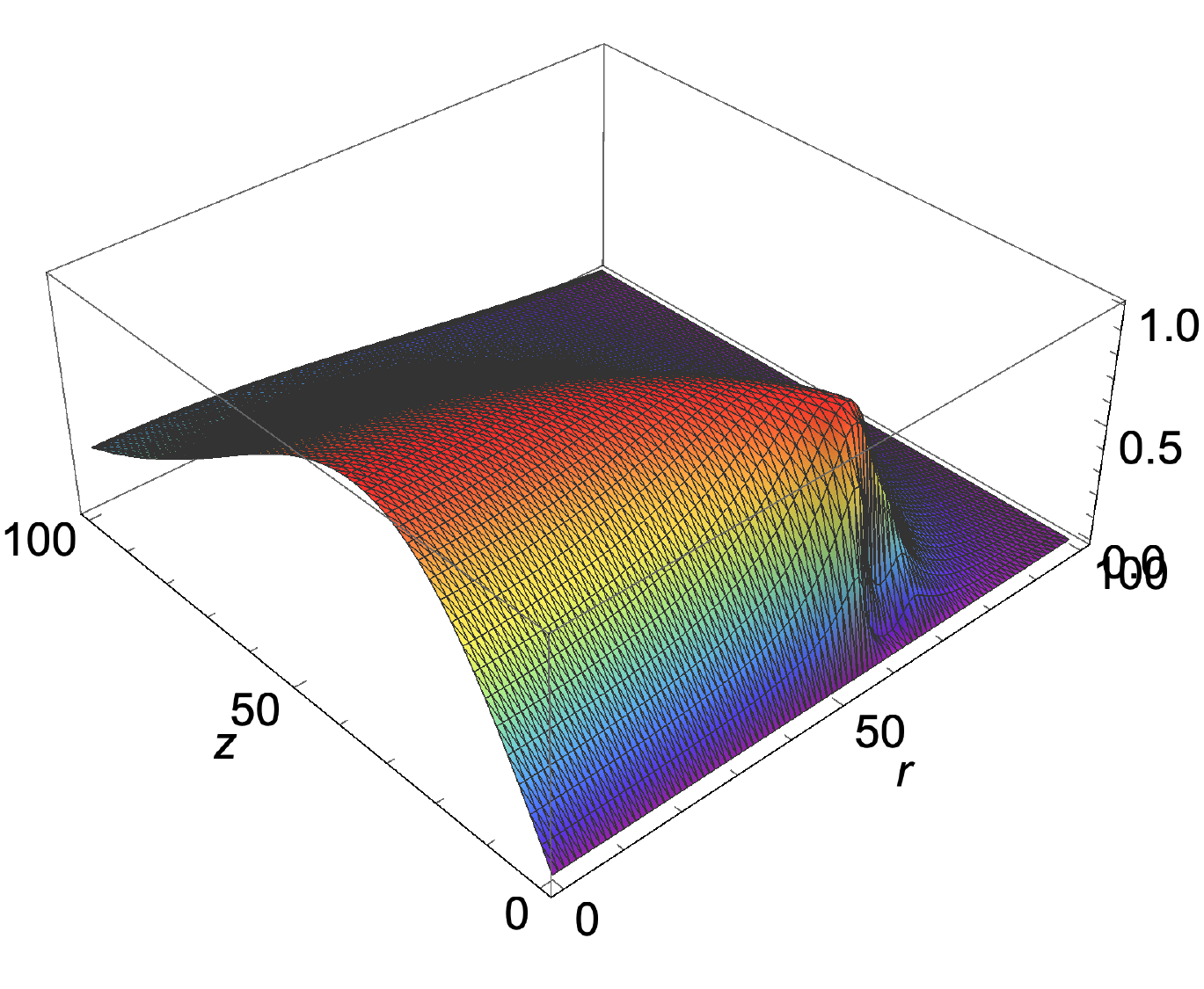}
\caption{$Y$ profile}
\end{subfigure}
\begin{subfigure}{.5\textwidth}
\centering
\includegraphics[width=0.9\linewidth]{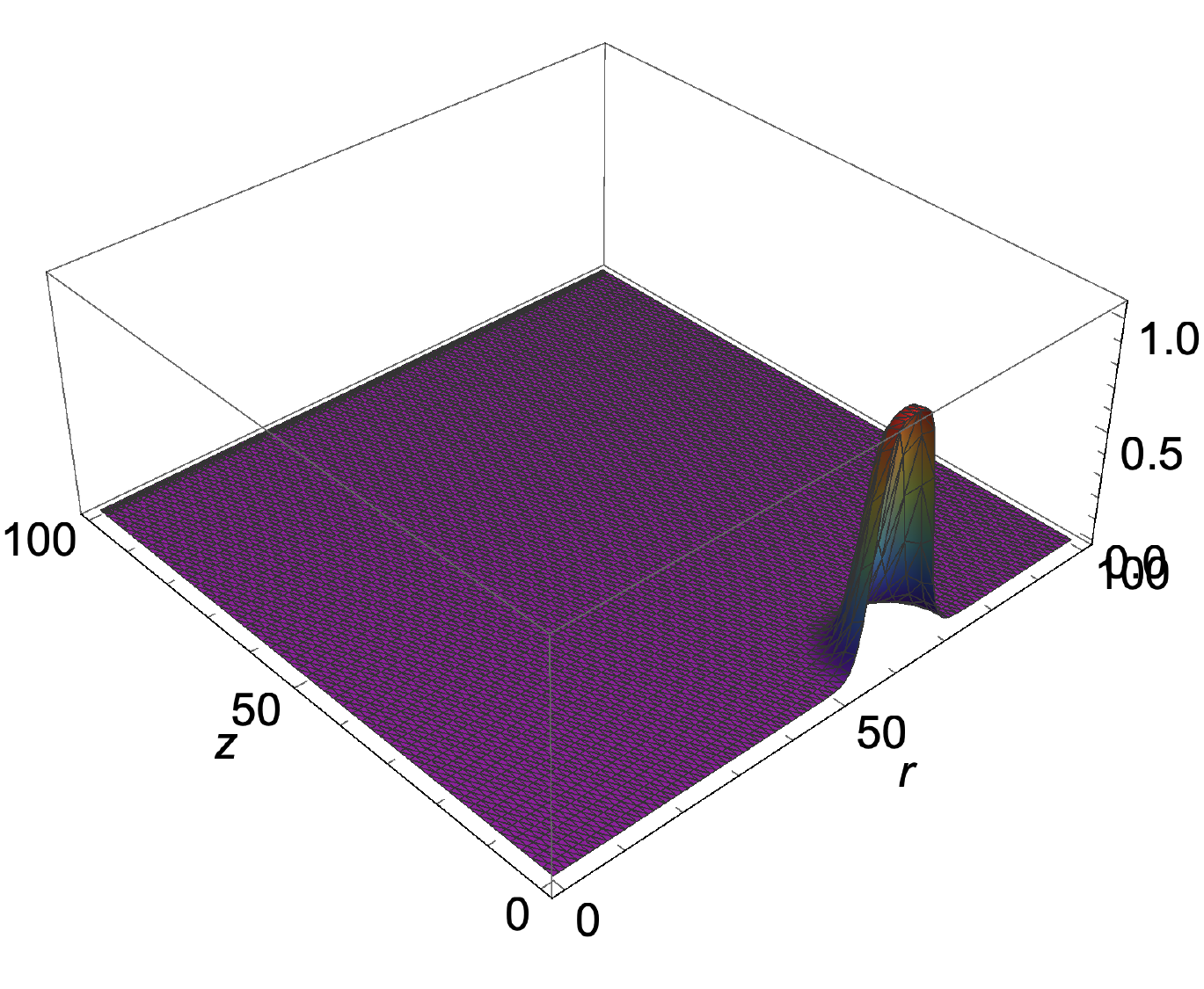}
\caption{$Z_1$ profile}
\end{subfigure}
\begin{subfigure}{.5\textwidth}
\centering
\includegraphics[width=0.9\linewidth]{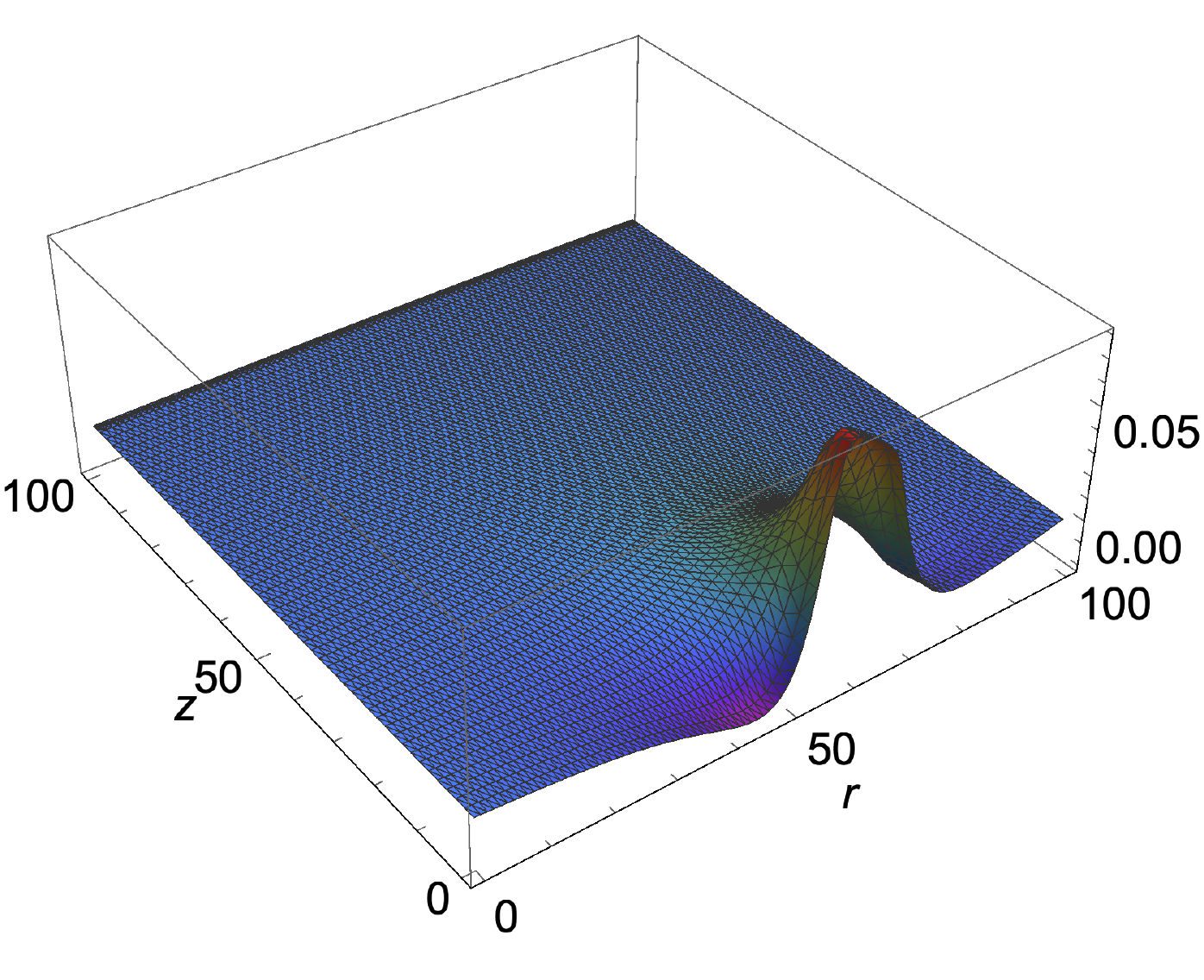}
\caption{$|\vec{B}|$}
\end{subfigure}
\begin{subfigure}{.5\textwidth}
\centering
\includegraphics[width=0.9\linewidth]{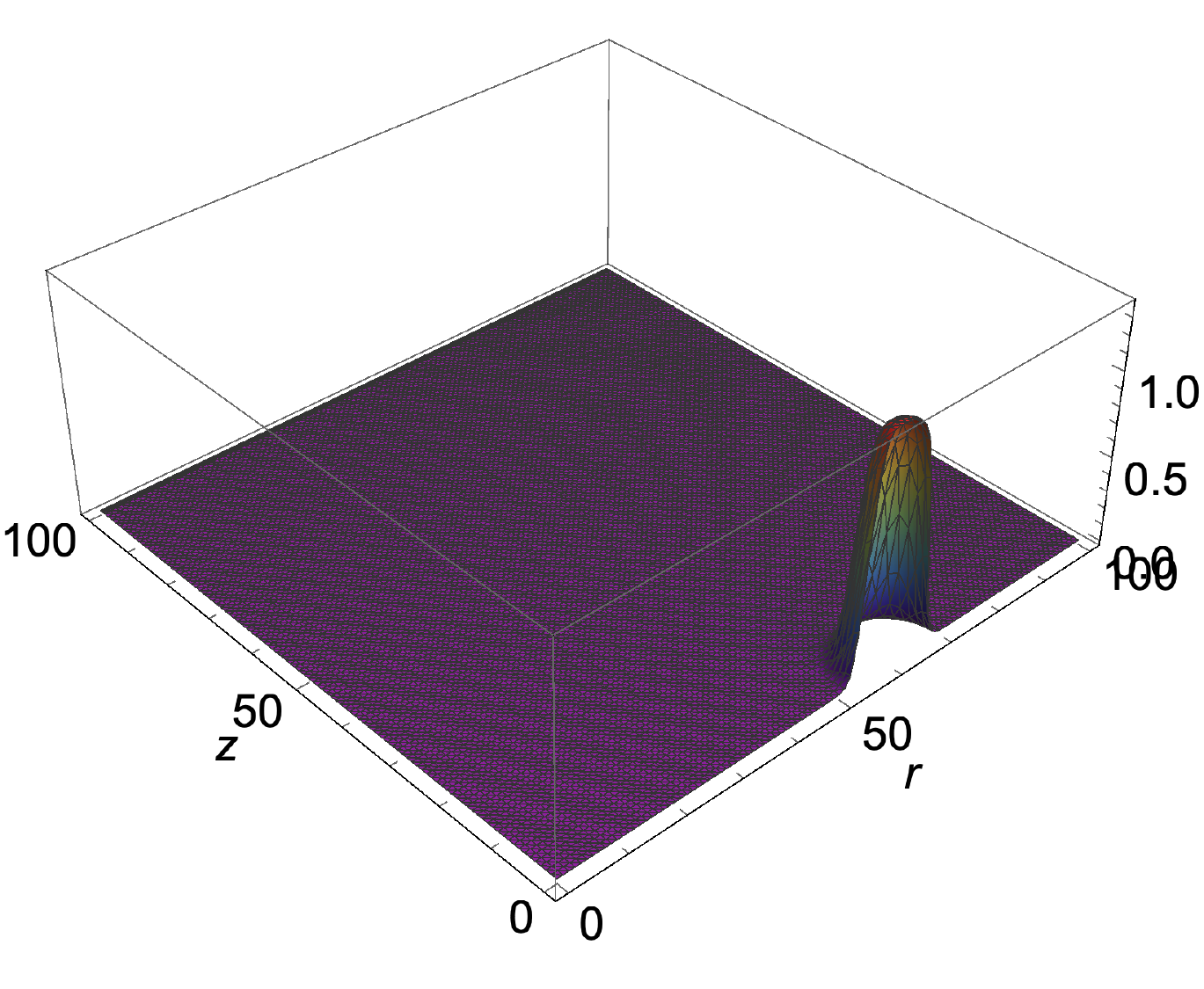}
\caption{Energy density profile}
\end{subfigure}
\caption{Vorton profiles for $X$, $Y$, and $Z_1$ components.  The magnetic field $|\vec{B}|=B_\phi=\partial_r A_z$ and energy density profile are also shown.  The parameters are $n=1$, $m = 40$, $\lambda_\phi = 4.5$, $\lambda_\sigma = 4.0$, $\gamma = 2.8$, $Q = 60000$, $g = 0.1$.  For this value of $Q$, we find $\omega = 0.576$.}%
\label{fig11}%
\end{figure}

%%% Revision request #20: We correct the designation of our approximation method.  This is gradient descent, not Newton-Raphson. %%%
An example of the solutions we obtained is  presented in Figure \ref{fig1}. 
The solutions were found using a standard gradient descent relaxation algorithm. As stated earlier the value of $Q$ was held constant, while the system in (\ref{FinalForm2}) was solved at each iteration with $\omega$ determined in (\ref{Charge1}). In Figure \ref{fig11} we show another solution which is a  representative of what happens near to the thin vorton limit.   %%%Revision request 21: The formula for the magnetic field is inserted for clarification. %%%

\begin{figure}[h!]. %%% Revision request #23: The decay plots have been updated with larger axes labels.  %%%
\centering
\includegraphics[width=0.55\linewidth]{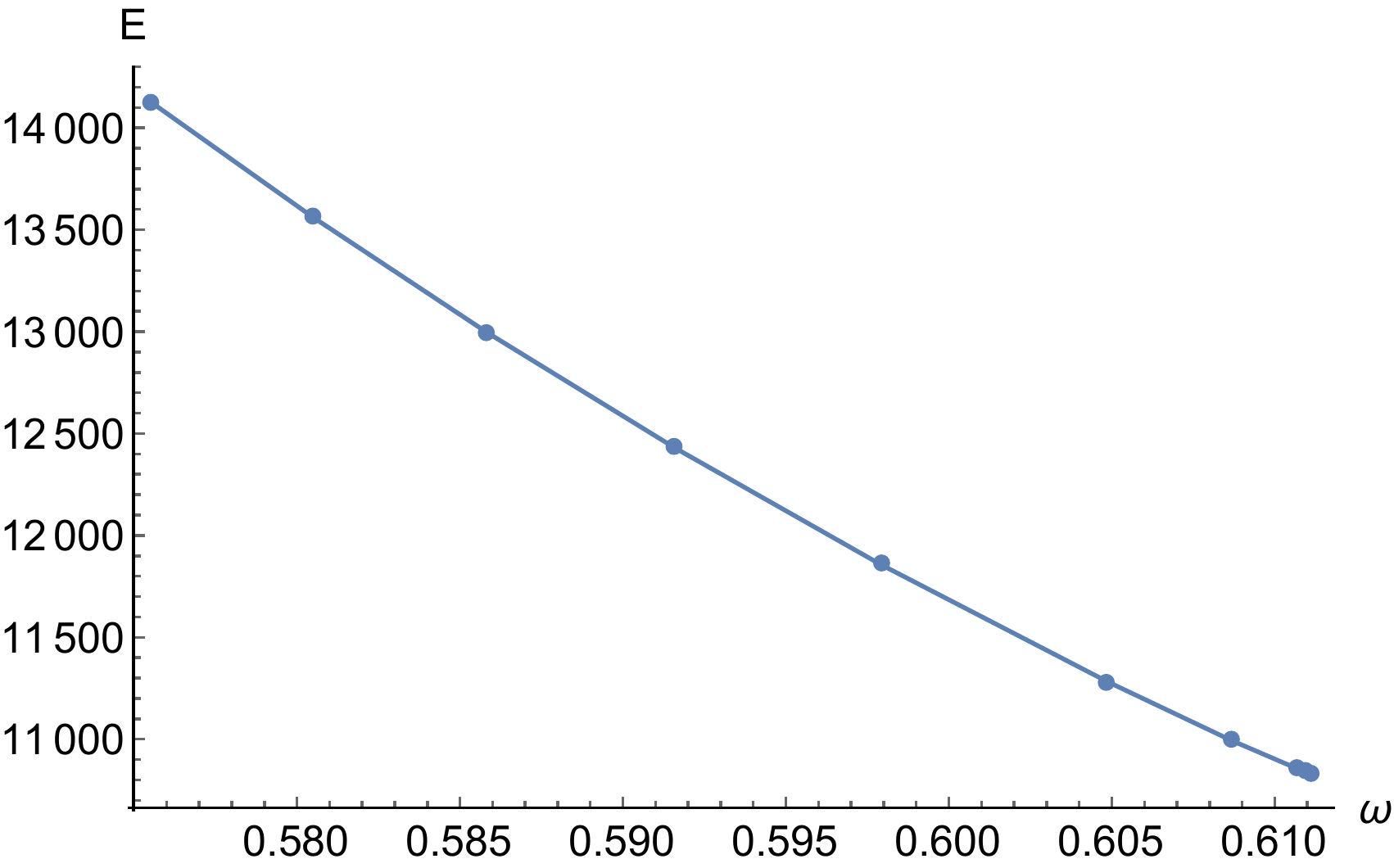}. 
\caption{Energy vs $\omega$ for the previous solution parameters at varying $Q$. The last point is where the vorton loses stability.}%
\label{fig2}%
\end{figure}

We investigated the dependence of the energy on $\omega$  which we present in Figure \ref{fig2}. Beyond the furthest point in $\omega$, which we call $\omega_{\rm crit}$ our solutions quickly lost convergence and decayed into a more stable solution.
%This was somewhat unexpected. 
%Of the U-shaped branch of solutions one expects only the left hand side to be stable \cite{Radu:2008pp} while the right hand branch to be unstable, and this is indeed what we found: a decreasing energy as a function of $\omega$ until, presumably, the increasing branch is found.  
\begin{figure}[h!]
\begin{subfigure}{.5\textwidth}
\centering
\includegraphics[width=0.8\linewidth]{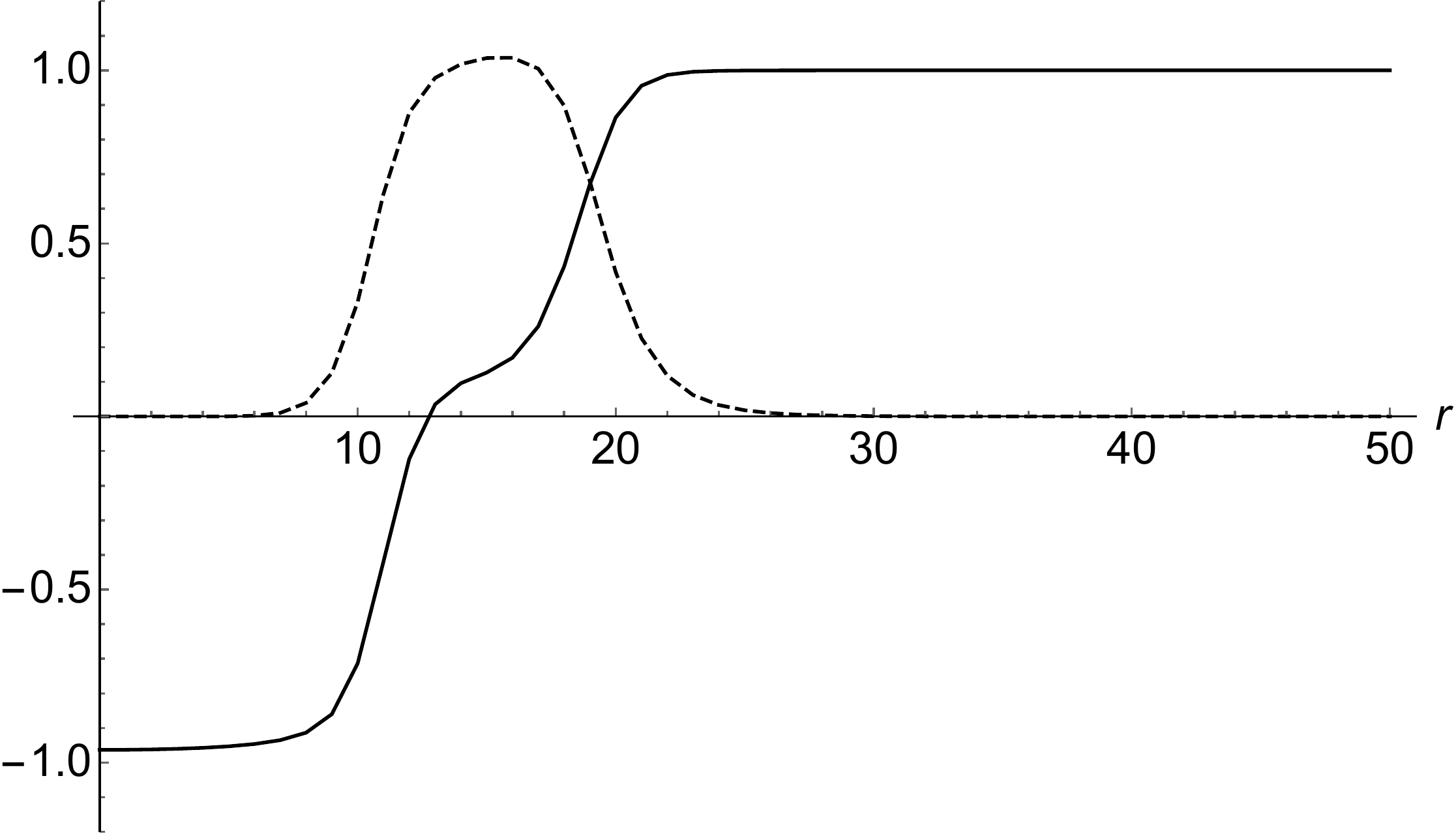}
\caption{$t_{\rm steps} = 0$}
\end{subfigure}
\begin{subfigure}{.5\textwidth}
\centering
\includegraphics[width=0.8\linewidth]{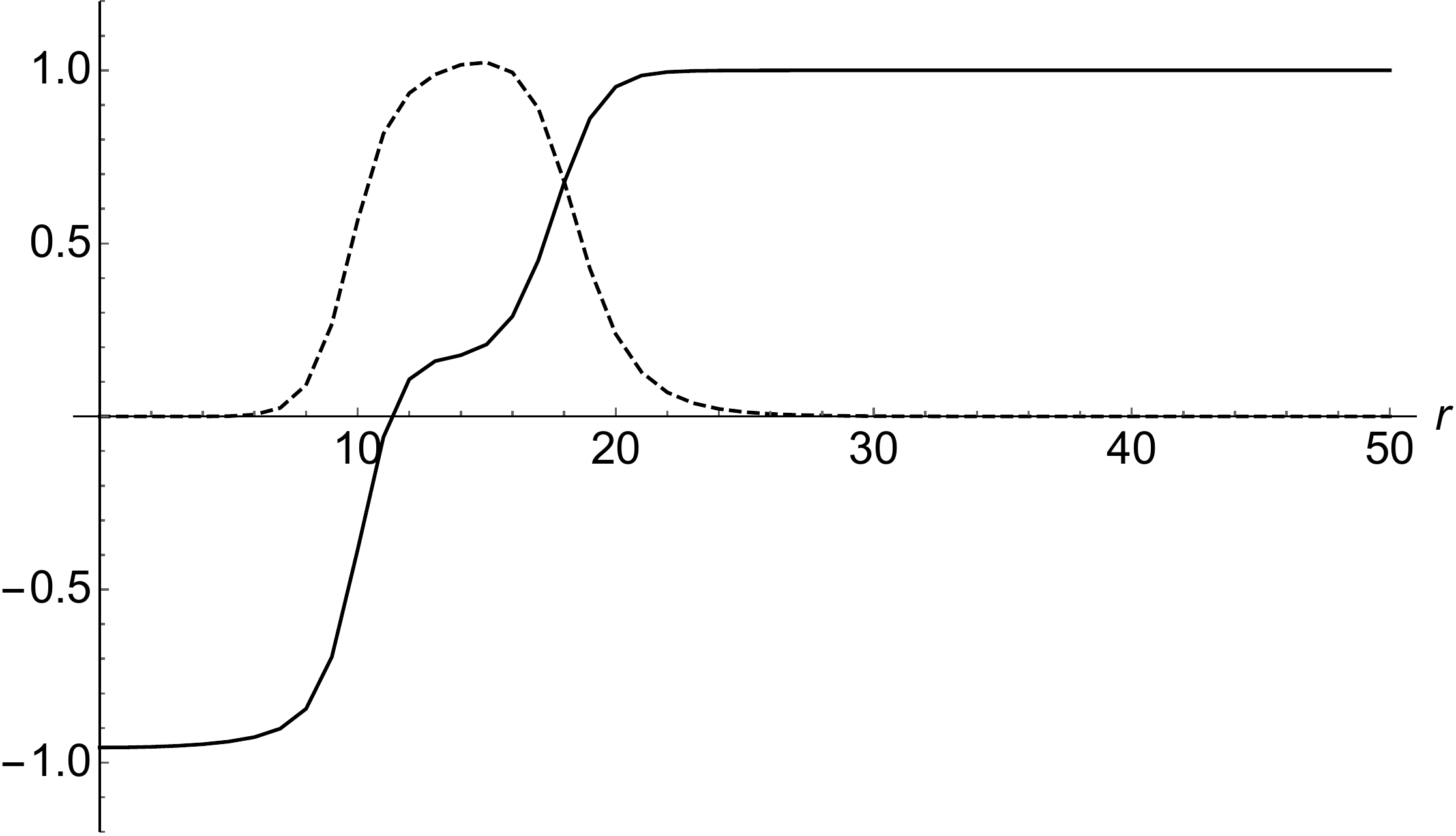}
\caption{$t_{\rm steps} = 7000$}
\end{subfigure}
\begin{subfigure}{.5\textwidth}
\centering
\includegraphics[width=0.8\linewidth]{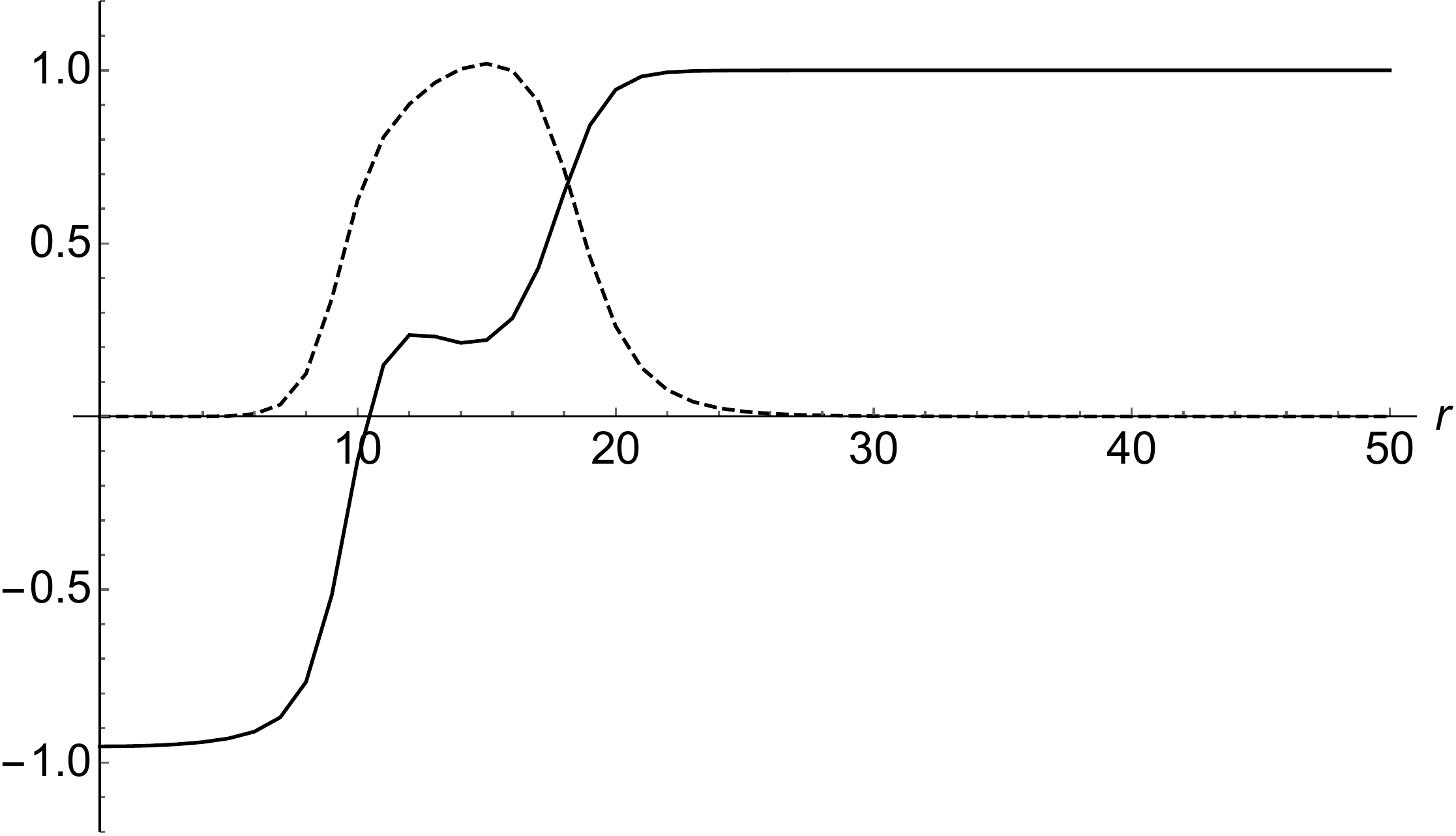}
\caption{$t_{\rm steps} = 9000$}
\end{subfigure}
\begin{subfigure}{.5\textwidth}
\centering
\includegraphics[width=0.8\linewidth]{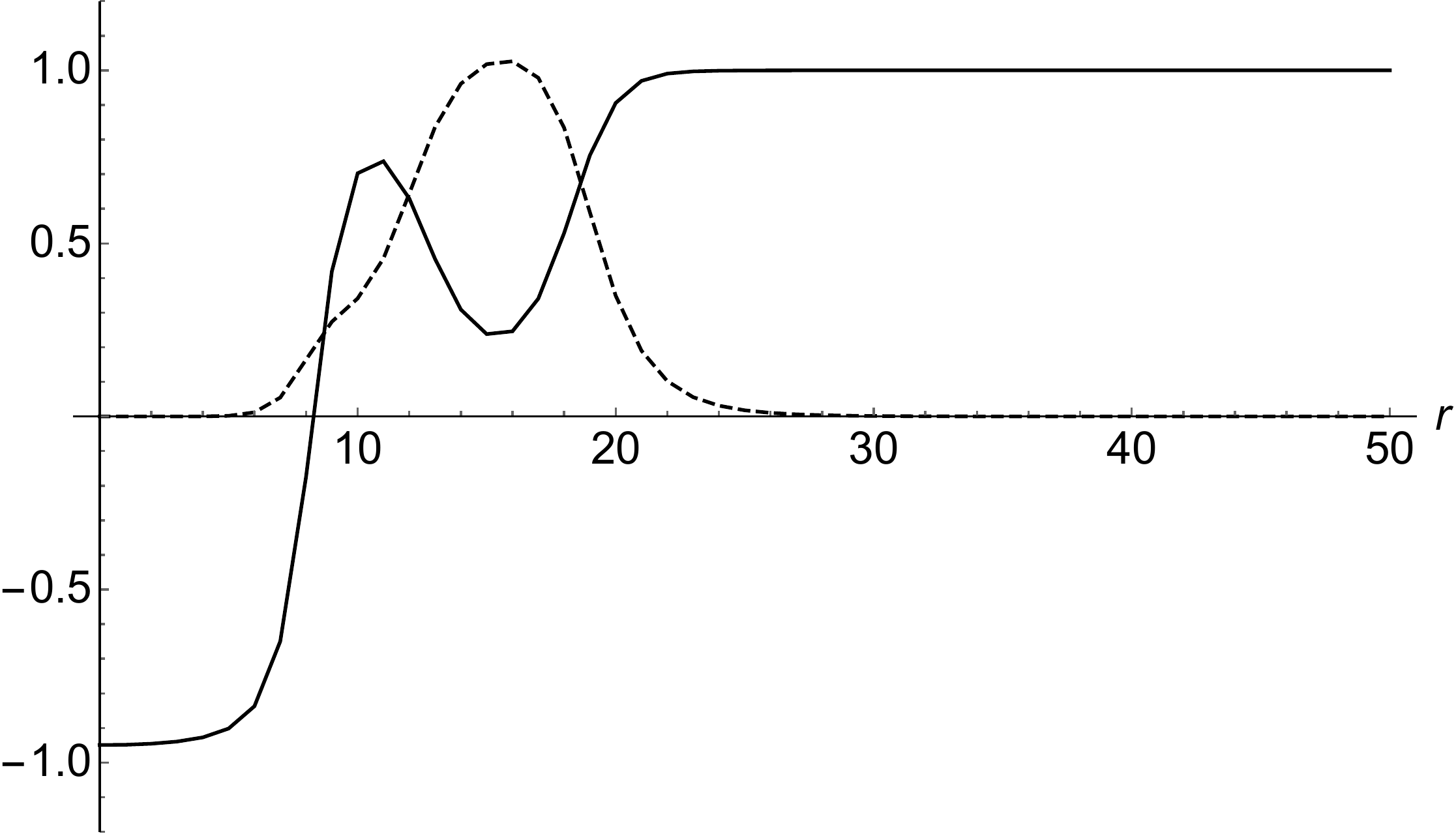}
\caption{$t_{\rm steps} = 9500$}
\end{subfigure}
\begin{subfigure}{.5\textwidth}
\centering
\includegraphics[width=0.8\linewidth]{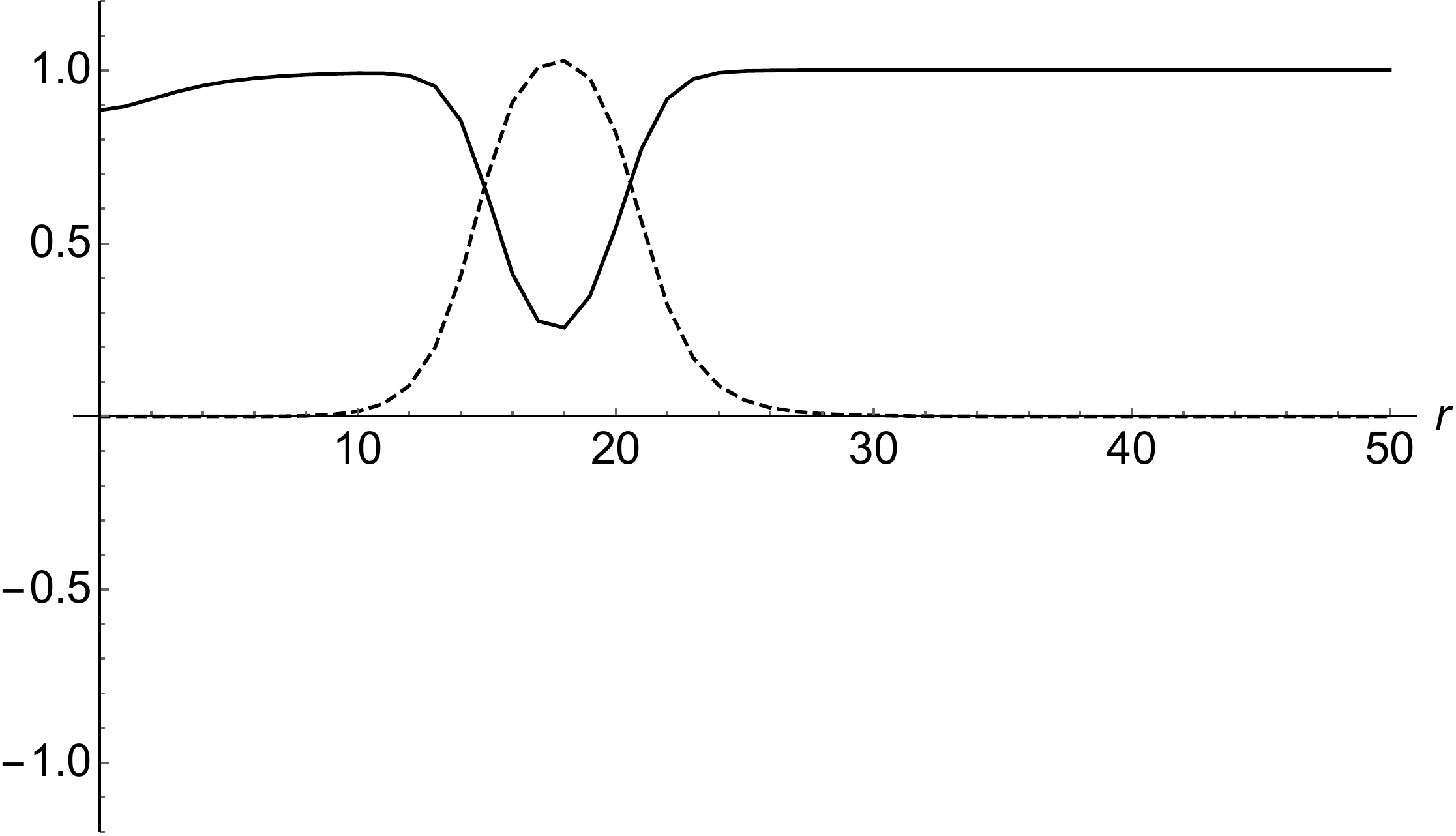}
\caption{$t_{\rm steps} = 10000$}
\end{subfigure}
\begin{subfigure}{.5\textwidth}
\centering
\includegraphics[width=0.8\linewidth]{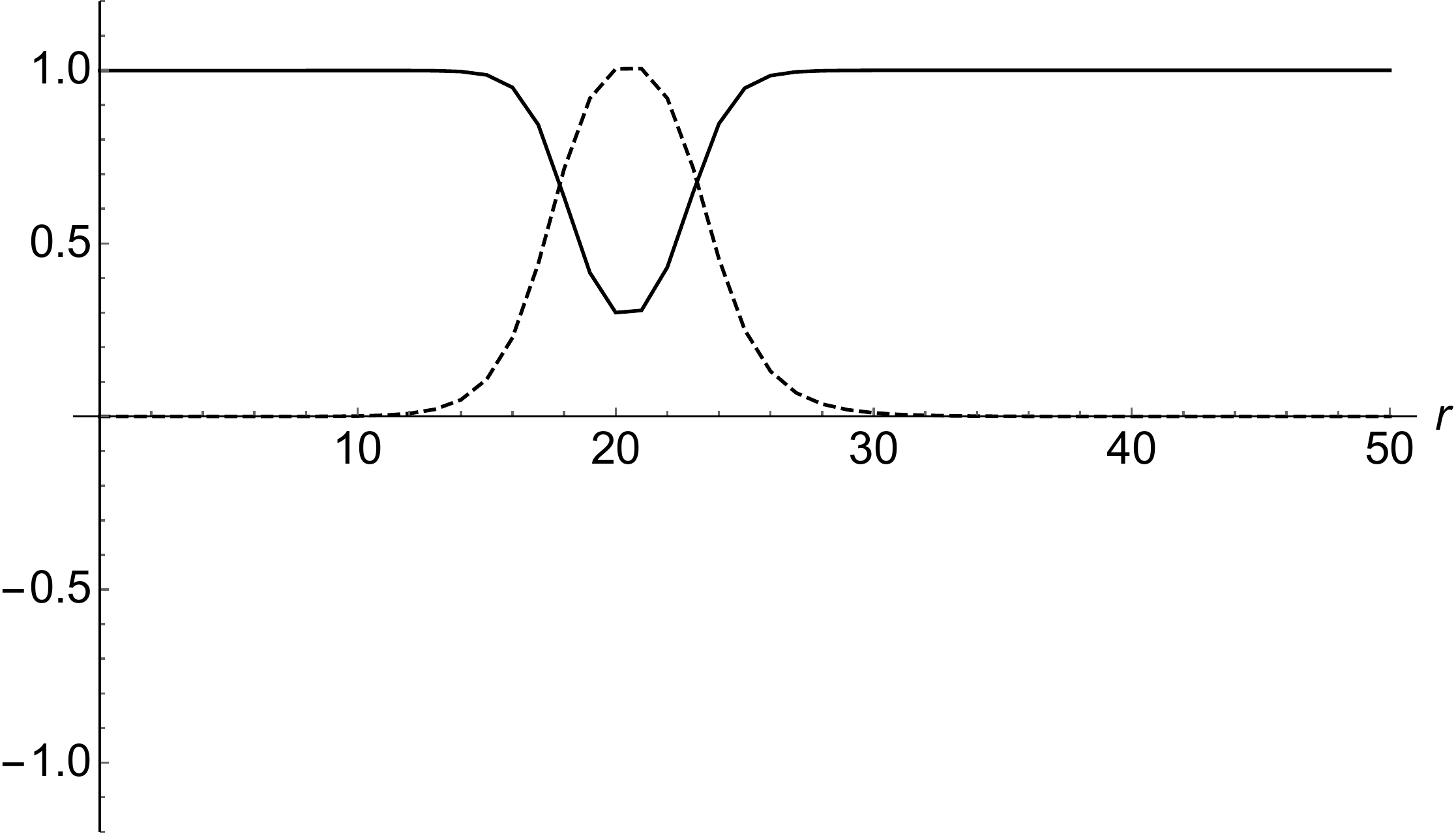}
\caption{$t_{\rm steps} = 11000$}
\end{subfigure}
\caption{$X$ (solid) and $Z$ (dashed) field configurations at $z = 0$ at various numbers of steps in the iteration procedure.  The procedure is relaxed for Q = 7000, and then set to Q = 6000, which is below the threshold for decay.  The parameters are $\lambda_\phi = 4.5$,  $\lambda_\sigma = 4.0$, $\eta = 1$, $\gamma = 2.8$, $m = 10$, and $g = 0.1$.}%
\label{fig22}%
\end{figure}
What precise channel of decay these vortons have in the right hand branch is a novel aspect of this paper and is detailed in Figure \ref{fig22}. The final state is a sort of spinning Q-balls with no vortex winding, i.e. $n=0$. The existence of this type of solution was suggested in  \cite{Radu:2008pp} but not found explicitly.\footnote{Q-rings in a related model have been recently discussed in  \cite{Loiko:2019gwk}.} Here we not only find them explicitly, but also show that they are energetically prefered to the vorton for certain $\omega$.   For all the parameter ranges we explored numerically, we could not avoid finding a certain critical $\omega$ after which a similar decay occurred.

\begin{figure}
\centering
\includegraphics[width=0.7\linewidth]{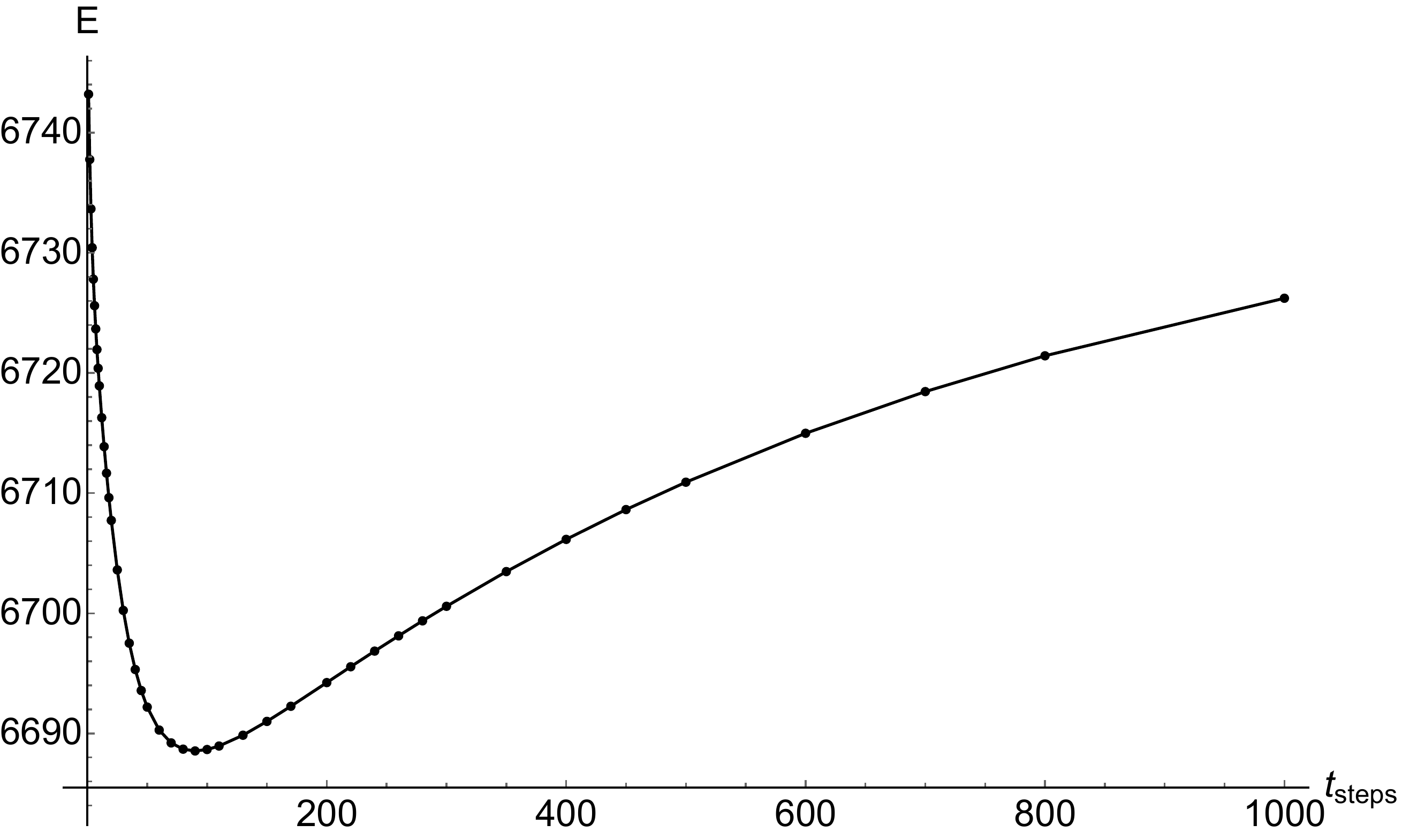}. %%%Revision request #23: Graphics text has been improved. %%%
\includegraphics[width=0.7\linewidth]{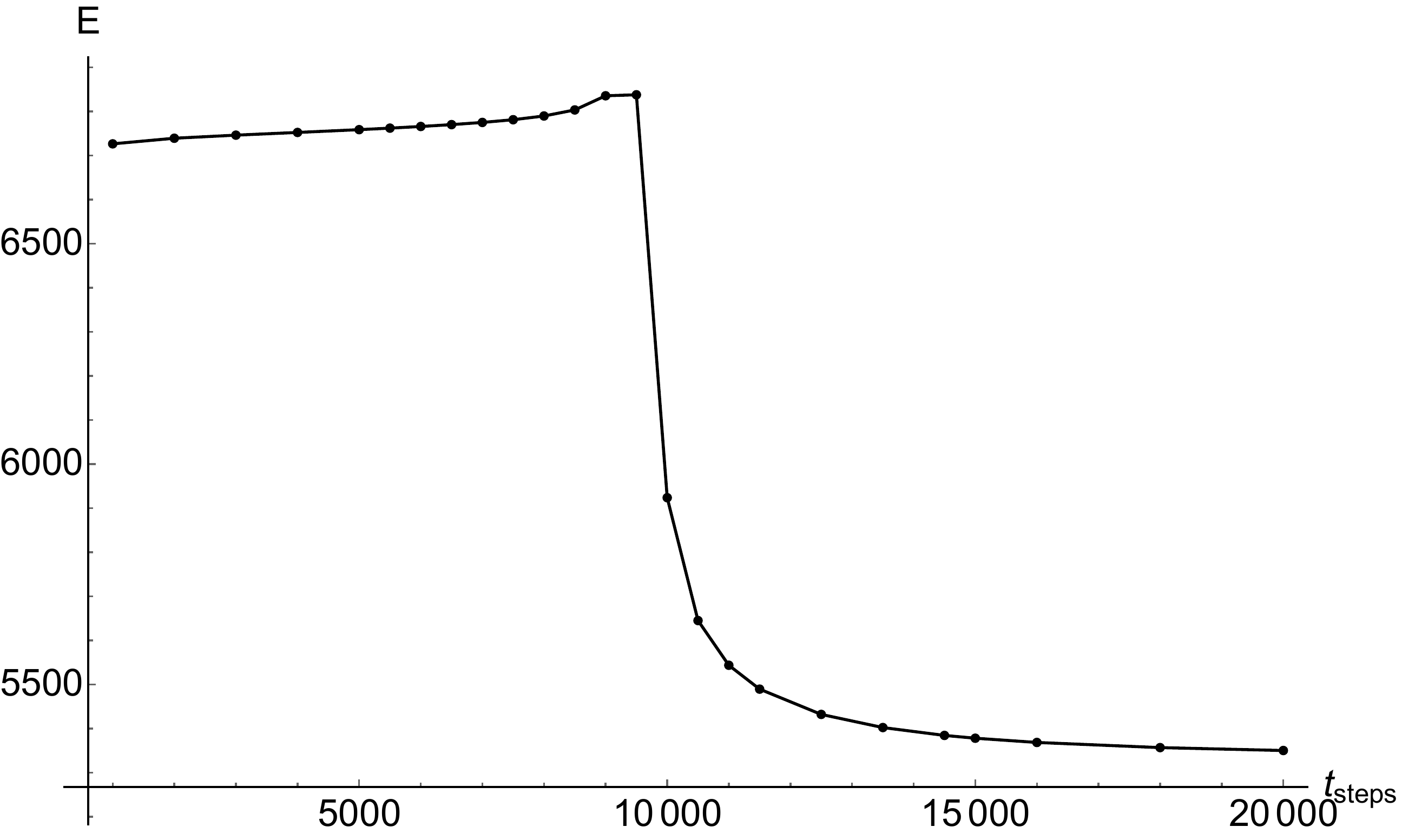}. %%%Revision request #23: Graphics text has been improved. %%%
\caption{We show the energy of the solutions as a function of iteration number corresponding to those shown in Figure \ref{fig22}.  The first graph illustrates the rapid descent to minimum energy and gradual increase in energy.  The second graph is taken over longer timescale and illustrates the metastability of the vorton just below threshold. There is a critical point at about $Q \sim 9000$ where the vorton rapidly decays to the rotating Q-ball configuration.}
\label{EvtStepsDecayPlot}
\end{figure}
%%% Revision request #22: We have extended the graphic in figure 6 out to t = 10000 to clarify the plateau behavior. %%%
\begin{figure}
\centering
\includegraphics[width = 0.7\linewidth]{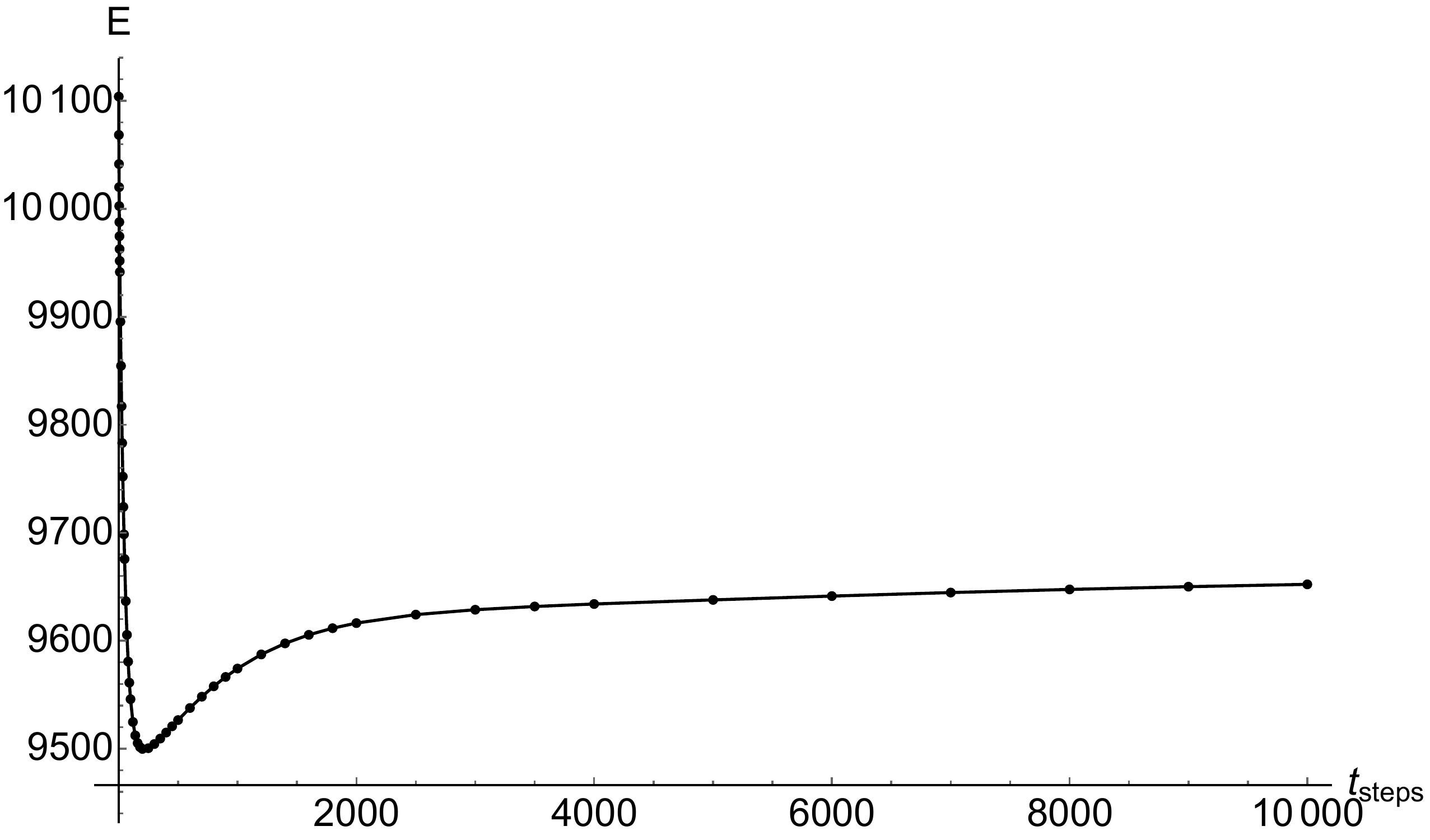} %%%Revision request #23: Graphics text has been improved. %%%
\caption{To illustrate the case for a vorton with Q above the threshold for decay, we show the energy as a function of iteration number in the procedure. For this case we allow the procedure to evolve a solution from $Q = 15000$ to $Q = 10000$, with all other parameters set to those in Figure \ref{fig22}.  We see a clear rapid descent to minimum and a slow rise to a constant plateau.}
\label{EvtStepsNoDecayPlot}
\end{figure}

To further illustrate the decay behavior of a vorton with charge $Q$ set below the critical value, we show the time step evolution of the energy of the configuration in Figure \ref{EvtStepsDecayPlot}.  We observe that the vorton configuration quickly descends to a minimum energy and begins to slowly rise until it reaches a critical point. At this point the vorton rapidly decays to the rotating Q-ball configuration. For comparison we have also included the time step evolution of our procedure for the case of a vorton with charge $Q$ set above the threshold at $Q = 10000$. In this case the energy quickly descend to a minimum and gradually rises to a plateau.

\begin{figure}[h]
\centering
\includegraphics[width=.7\linewidth]{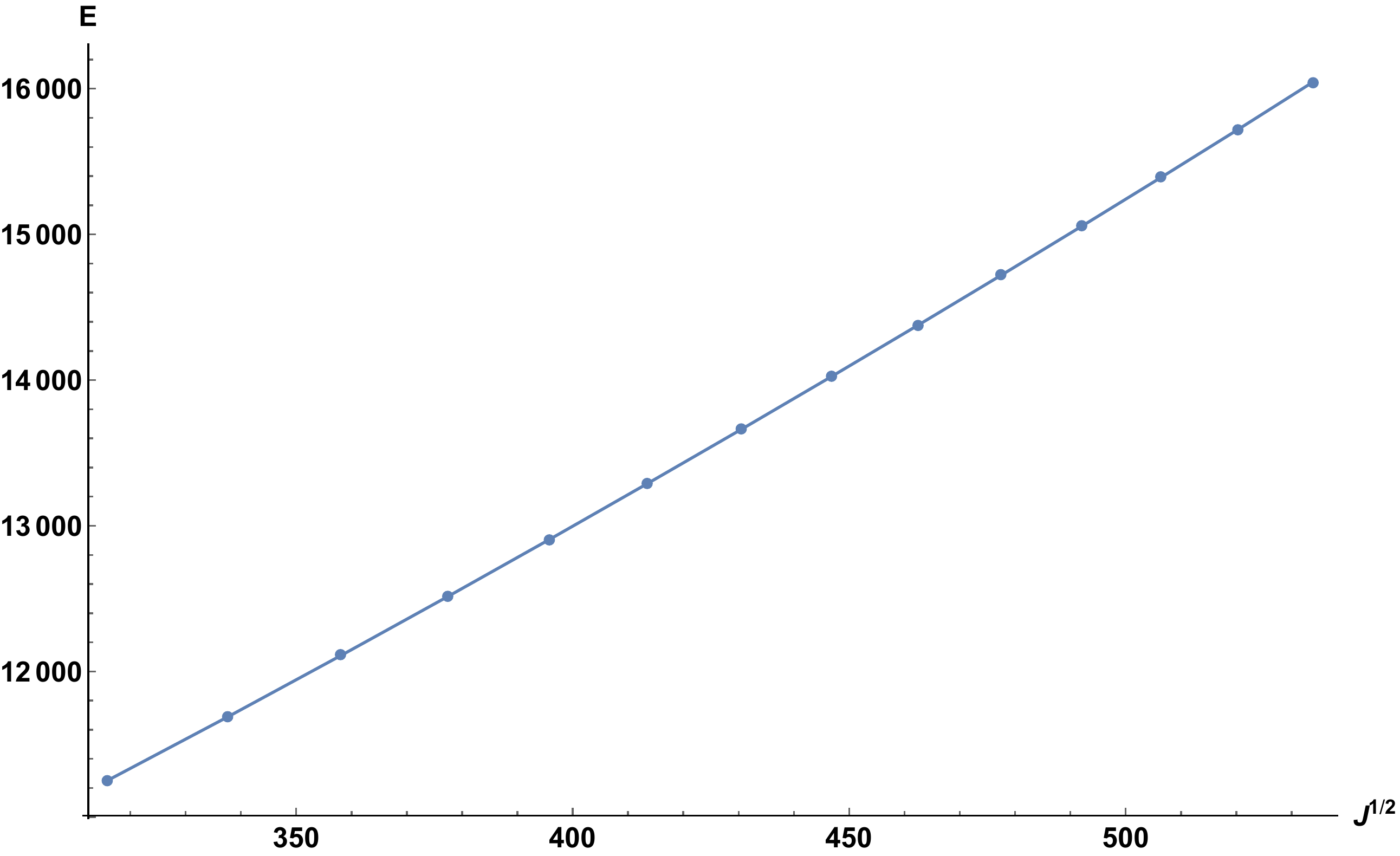}
\caption{Energy vs $\sqrt{J_3}$ for a vorton close to the critical $\omega$ for $Q=14250$.
%, (a), and for a vorton far from it, (b).
}%
\label{fig44}%
\end{figure}

In Figure \ref{fig44} we present a plot of the energy $E$ vs $J_3^{1/2}$ in order to observe the Regge trajectory one would expect in the thin vorton limit  (\ref{regge1}). We plot the values close to $\omega_{\rm crit}$. We find that the solutions satisfy this kind of trajectory closely, but that some degree of curvature exists, which %, as can be seen from Figure \ref{fig55}, which shows a plot of the gradients of the previous graph.
can be traced to the fact that vorton is not exactly thin and moreover we can never reach the optimal value of $\omega$, the best that we can do is to compare the various solutions close to the critical value.

\begin{figure}[h!]
\centering
\includegraphics[width=0.4\linewidth]{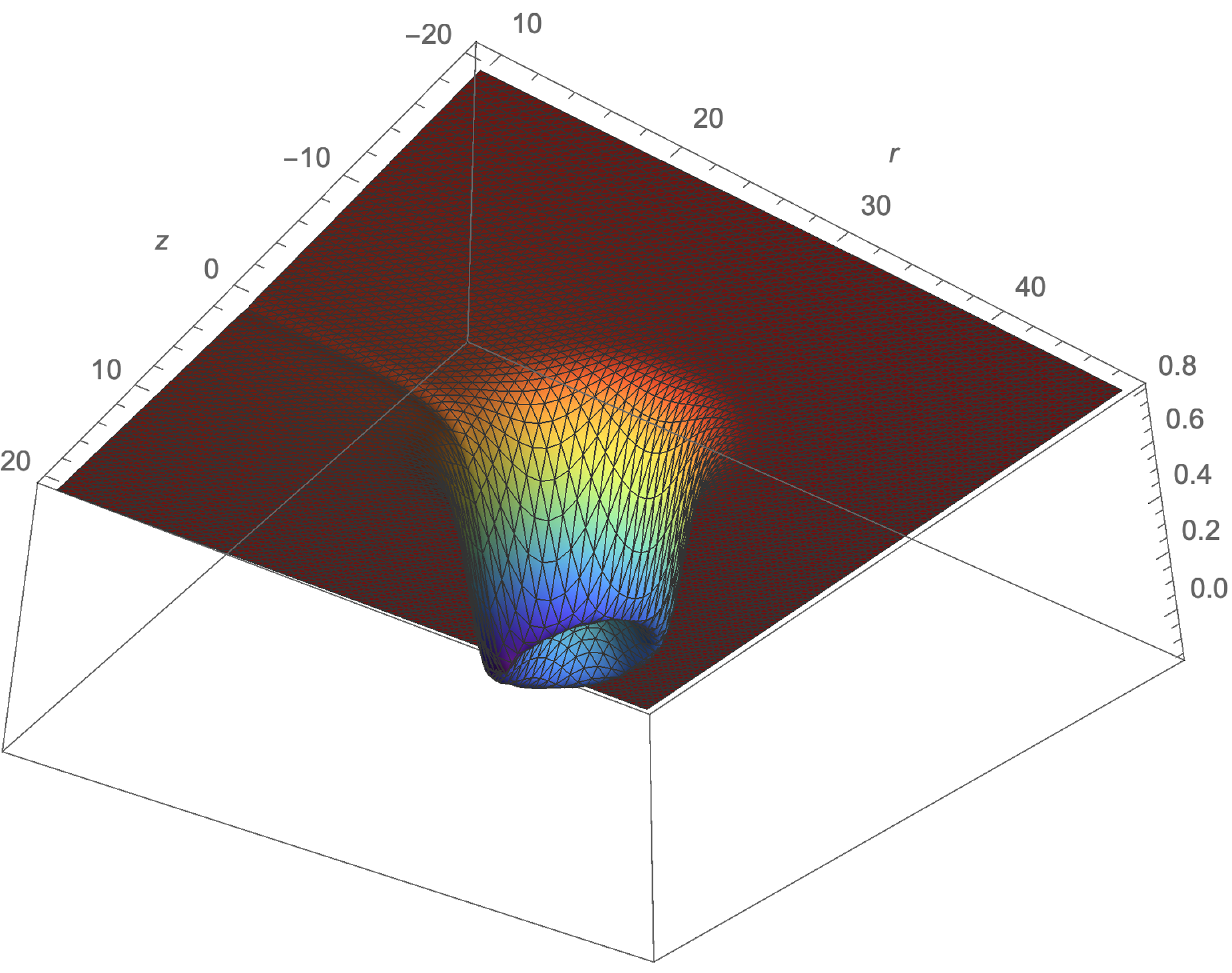} \qquad 
\includegraphics[width=0.5\linewidth]{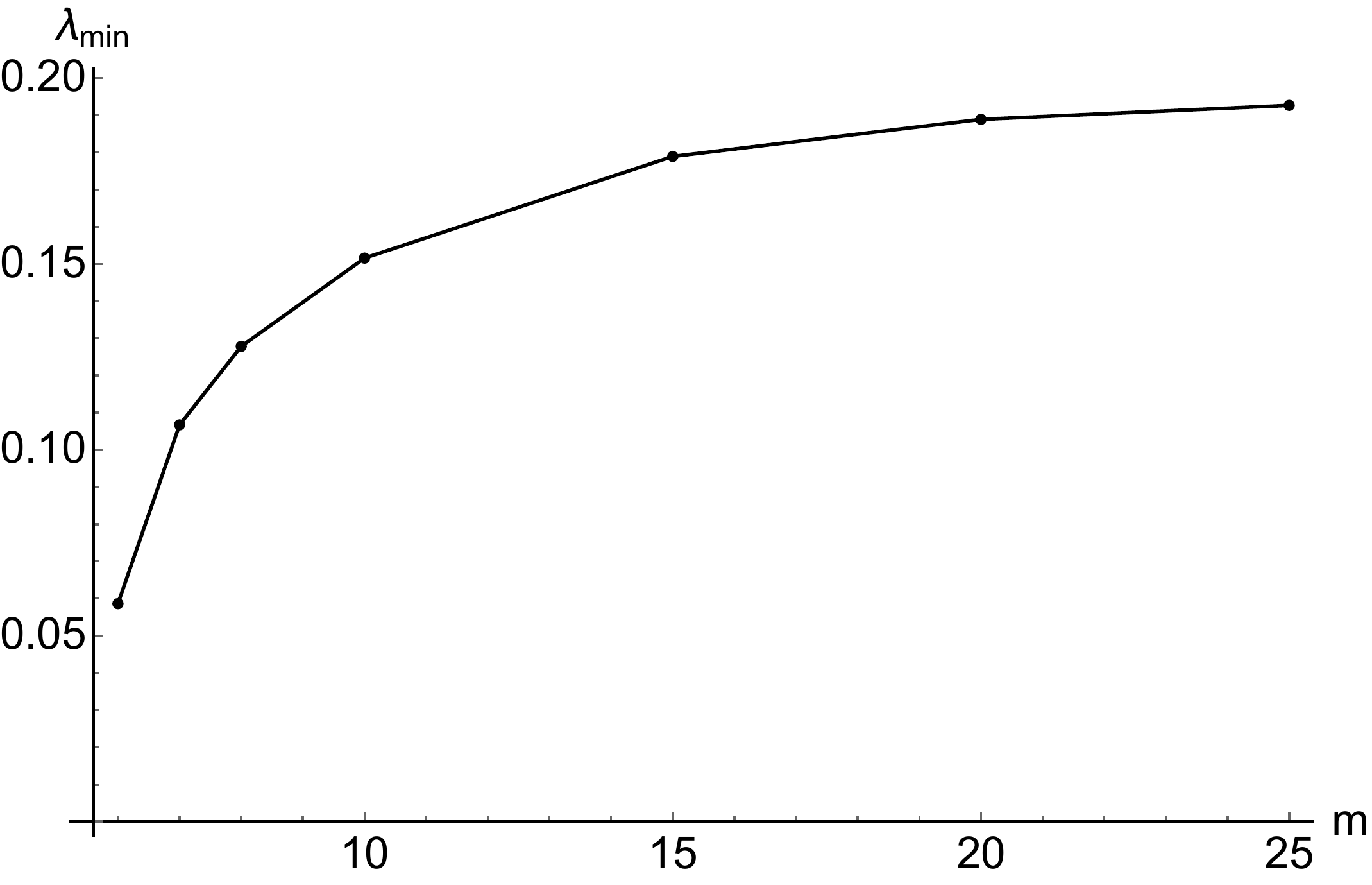}
\caption{Left panel: the effective potential for $Z_2$ perturbations following from (\ref{EffectivePotential}). Right panel: lowest eigenvalue of linearised Schrodinger problem for $Z_2$ as a function of $m$ for the solution parameters shown in Fig. 1 and $Q=15000$.}%
\label{Z2}
\end{figure}

All the plots we presented can be considered as solutions of the Abelian model, as well as embedding into the non-Abelian one since the non-Abelian field is always confined to an equatorial plane of internal space. 
Following the numerical construction of the vorton in the non-Abelian model we ask whether small perturbations of $\chi^i$ out of the $i = \{1,2\}$ plane will lead to solutions outside the equatorial plane.  To clarify, we are not evaluating the general stability of our solutions in a full 3 dimensional time dependent case. Here, we are just interested in the stability of $U(1) \times SO(3)$ axially symmetric vortons to axially symmetric perturbations on the $\chi^3$ field. 
%This would indicate a channel of convergence where solutions would present a non-zero $Z_2$ and therefore be truly non-Abelian in nature.
To determine the stability of the vorton under small perturbations of the $\chi^3$ field we consider the effective potential for small $\chi^3$ perturbations in the equations of motion.  Following from (\ref{FinalForm2}) we have
\begin{equation}
V_{\rm eff}(r,z) = \frac{\lambda_\chi}{2}\left(Z_1^2 -\eta_\chi^2\right) + \gamma \left( X^2+Y^2\right).
\label{EffectivePotential}
\end{equation}
The profile for this potential is shown in Figure \ref{Z2} lefthand side.  This potential is mostly positive for the parameters we have chosen.  However, small regions in the potential well indicate possible instabilities that we must check in the equations of motion for $Z_2$.
To evaluate the stability of the vorton to small perturbations of $Z_2$ in the potential (\ref{EffectivePotential}) we must solve the eigenvalue problem
\begin{equation}
\left(-\nabla^2 + V_{\rm eff}\right) Z_2 = \omega Z_2,
\end{equation}
and determine if negative eigenvalues exist, indicating instabilities of the $U(1) \times SO(3)$ vorton. We perform the eigenvalue analysis for several values of $m$ in Figure \ref{Z2} right hand side.  We find that for higher values of $m$ the vorton has stronger stability towards $Z_2$ perturbations. We could not find any region of parameter space where an instability towards forming $Z_2$ exists. This indicates that these fields will preferentially want to settle into the internal equatorial plane.
Thus for all the solutions that converged, we were not able to find any for which $Z_2\neq 0 $, which were expected from the  analysis of the previous section. 
 All the other types of solutions cannot be reached, at the moment, by our numerical algorithm.

\section{Conclusion}
\label{cinque}

In this paper we studied vorton solution in an Abelian model, which is a global version  of the Witten superconducting string model, and a non-Abelian generalization where a triplet of fields transforming under $SO(3)$ condenses in the core of the vortex. Numerical solutions for vortons in the Abelian model have been found previously. We have recovered those solutions with our technique, finding also new results. First we have been able to perform the numerics for quite large values of parameters, finding qualitative agreement with certain properties expected from the thin vorton approximation, for example
Regge type behaviour. We also discovered an instability channel of vorton solutions for $\omega > \omega_{\rm crit}$, explicitly showing their decay into Q-ring solutions.

We also studied the inclusion of non-Abelian degrees of freedom to vorton solutions, and the stabilization of these by non-Abelian currents.  For all the numerical runs we have performed we always found solutions where the internal condensate would stay in the equator of the inner sphere, thus rendering the solution essentially an embedding of the one obtained in the Abelian case. Nevertheless  the choice of the equatorial plane can be arbitrary, as the generator of the $SO(3)$ current that stabilizes the vorton, and so the soliton has an internal non-Abelian degree of freedom.   This research would also be interesting in other types of non-Abelian vortex setups \cite{Hanany:2003hp,Auzzi:2003fs}. However, these setups are more complicated due to the presence of non-Abelian gauge fields. In this sense, the present model is a good toy model for this analysis.

\section*{Acknowledgments}

We thank the referee for useful comments.
The work of G.T. is funded by a Fondecyt grant number 11160010.  The work of A.P. was funded in part by the NSERC Discovery Grant. The work of S.B. is  supported by the INFN special project grant ``GAST (Gauge and String Theory)''.

\end{document}